\documentclass[11pt]{article}

\pdfoutput=1
\usepackage{graphicx}
\usepackage{amssymb}
\usepackage{amsmath}
\usepackage{bm}
\usepackage{latexsym}
\usepackage{textcomp}
\usepackage{color}
\usepackage{slashed}
\usepackage{xcolor,colortbl}

\newcommand{\mc}[2]{\multicolumn{#1}{c}{#2}}
\definecolor{Gray}{gray}{0.9}
\newcolumntype{a}{>{\columncolor{Gray}}c}

\usepackage[a4paper]{geometry}
\makeatother

\begin{document}


\title{On the covariant formalism of the effective field theory of gravity and leading order corrections}
\author{Alessandro Codello\thanks{codello@cp3-origins.net} \,and Rajeev Kumar Jain\thanks{jain@cp3.sdu.dk}\\
\emph{CP$^3$-Origins, Centre for Cosmology and Particle Physics Phenomenology}\\
\emph{University of Southern Denmark}\\
\emph{Campusvej 55, 5230 Odense M}\\
\emph{ Denmark}}
%
\date{}

\maketitle

\begin{abstract}
We construct the covariant effective field theory of gravity as an expansion in inverse powers of the Planck mass, 
identifying the leading and next--to--leading quantum corrections.
We determine the form of the effective action for the cases of pure gravity with cosmological constant as well as gravity coupled to matter.
By means of heat kernel methods we renormalize and compute the leading quantum corrections to quadratic order in a curvature expansion.
The final effective action in our covariant formalism is generally non--local and can be readily used to understand the phenomenology on different spacetimes.
In particular, we point out that on curved backgrounds the observable leading quantum gravitational effects are less suppressed than on Minkowski spacetime.
%
\end{abstract}


\section{Introduction}

In recent years the fact that quantum gravity can be treated as an effective field theory (EFT) has become an established fact \cite{Donoghue:1993eb,Donoghue:1994dn,Burgess:2003jk,Donoghue:2012zc,Donoghue:2015hwa}.
The modern perspective is that even the standard model of particle physics, originally constructed under the guiding principle of perturbative renormalizability, must be seen as an EFT \cite{Donoghue:1992dd,Pich:1998xt,Burgess:2007pt,Weinberg:2009bg}. The same applies to QED, theory that claims the best comparison between theory and experiment, together with many other successful quantum field theories.  Thus, in complete opposition to what has been thought for a long time, gravity can be treated as a perturbative (effective) quantum field theory alongside with the other fundamental forces.
It is now also recognized that there is no fundamental inconsistency between general relativity and quantum mechanics, at least at low energies: the EFT of gravity is a perfectly well defined and predictive quantum theory of gravity. While the standard model breaks down at a given UV scale, the EFT of gravity will break down at the Planck scale.
%
%
Any UV completion of gravity will have to reproduce the EFT predictions at sufficiently low energies. This is one reason why we should work out the EFT of gravity.
On the other side, as the example of QCD has shown clearly, even if we know the UV completion of a theory, so that all physical questions are mathematically well posed, it can still be extremely difficult to work out the low energy spectrum and the effective action. Thus even if we knew today the fundamental description of gravity at the Planck scale, we will probably still need to resort to EFT techniques to work out the accessible phenomenology.

To date, the EFT of gravity has been developed on a Minkowski background, using Feynman diagram techniques \cite{Donoghue:1994dn,Akhundov:1996jd,Khriplovich:2002bt,BjerrumBohr:2002kt}, or their modern versions \cite{Dunbar:1994bn,Dunbar:1995ed,Bern:2002kj,Bjerrum-Bohr:2013bxa}, and power counting arguments \cite{Donoghue:1996mt}.
The paramount result is the calculation of the leading quantum gravity corrections to the Newtonian potential \cite{Donoghue:1993eb}.
This is probably the most reliable result we have in quantum gravity so far,
but these corrections are very small and a possible direct observation of their effects is for the moment quite remote.
This fact in turn explains why Einstein's theory of general relativity (GR) is so successful, and predicts that it will still be for a broad range of energy scales.
The smallness of the quantum effects is the result of the wide separation between the scales where observations are made and the Planck scale.
This has lead to state, somehow paradoxically, that the EFT of gravity is the best perturbative quantum field theory we have \cite{Donoghue:2012zc}.
Recently, the quantum gravitational corrections to the bending of light by the sun have been computed \cite{Bjerrum-Bohr:2014zsa} and the first 
corrections to the black holes metrics are also known \cite{BjerrumBohr:2002ks,Kirilin:2006en}.

It is probable that if we will ever observe quantum gravitational effects,
these will be those already contained in the EFT of gravity,
despite our knowledge or ignorance about the underlying fundamental theory.
Since these EFT corrections are well defined, finite and computable with available techniques, or with foreseeable development of thereof, it becomes extremely important to fully determine them and to successively apply the resulting theory to any possible physical situation wherein an enhancement of such quantum effects could take place. 
Cosmology is probably the best setting to look for such quantum imprints, but to be able to consistently analyze the EFT corrections on a Friedmann--Robertson--Walker (FRW) background, one first needs to compute the effective action on an arbitrary background and then look for consistent solutions to the effective equations of motion (EOM).
In order to perform this task, we need to develop a {\it covariant} EFT of gravity. This is the primary purpose of this work, which focuses on the covariant formalism, the effective action and its curvature expansion.

We will start by showing how the EFT of gravity can be constructed as a saddle point, or a loop expansion in inverse powers of the UV scale, i.e. the Planck mass, within the covariant formulation of quantum gravity based on the background field method formalism. We will then identify the classical theory (CT), the leading order (LO) quantum corrections and also present the form of the next--to--LO (NLO) corrections. We also outline some features of the next--to--next--to--LO (NNLO) corrections in the case of gravity coupled to matter.
As mentioned earlier, it is probable that only the LO corrections will ever have a chance of being observed and so we will particularly focus on them, expressing all the pertaining details.
Unfortunately, in four dimensions, it is not known how to completely compute the functional traces involved in the LO corrections, so we will proceed by presenting the application of the heat kernel expansion, both in its local and non--local forms, to their evaluation. At present, heat kernel methods represent the state of the art techniques to tackle these kind of contributions. We will use the local expansion to discuss carefully the UV divergencies and the related renormalization, which, as in any EFT, is performed order by order at the cost of an input phenomenological parameter for any divergent coupling. Then we will introduce the non--local expansion which can be used to evaluate the LO contributions to second order in a curvature expansion.
We will discuss both the cases of a pure gravity EFT and its coupling to matter. In this paper we will restrict to LO corrections for the latter case for simplicity and leave for a future study the evaluation of those NLO and NNLO contributions leading to the covariant effective action that, when evaluated on a Minkowski background, gives the corrections to Newton's potential obtained by the flat space methods discussed earlier.
%
One particular reason behind our analysis being interesting is that it allows us to make a rational organization of various quantum gravity results that have been obtained over the years. For instance, in the EFT framework, UV divergencies can be turned into a physical result describing how the phenomenological parameters depend on the reference scale. Also the conformal anomaly is naturally included, as it induces the presence of a particular set of finite terms in the LO effective action, but since these terms {\it only} start to appear at third order in curvatures they will not be further discussed in this paper and are left to future analysis. 

The paper is organized as follows: in section \ref{section_one} we develop the covariant version of the EFT of gravity for the case of pure gravity.
In section \ref{section_matter} we include matter.
We then discuss renormalization and local finite terms in section \ref{section_renor} while in section \ref{section_nlna} we describe the non--local finite terms that can be inferred from the UV divergencies. In section \ref{section_curvature} we expose the curvature expansion to second order and obtain the final form for our LO effective action. Finally in section \ref{conclusions} we summarize our results and outline their implications. In the appendix we shortly review the local and non--local heat kernel methods used in the computation of the LO effective action.

\section{EFT of gravity}\label{section_one}

We don't know the UV completion of gravity,
but whatever degrees of freedom are associated with it and whichever symmetry
characterizes it, at energies much lower than the Planck scale,
the theory is broken to a vacuum invariant under diffeomorphisms.
This vacuum is the background physical geometry, described by the metric $g_{\mu\nu}$,
and the remnant quantum fluctuations are typically suppressed
by inverse powers of the Planck scale and can be treated in an effective manner.
Thus we can construct an EFT of gravity by quantizing metric fluctuations $h_{\mu\nu}$ around the background metric,
\begin{equation}
g_{\mu\nu}\rightarrow g_{\mu\nu}+\frac{1}{M}h_{\mu\nu}\,,
\label{exp_metric}
\end{equation}
where the scale $M$, defined by
\begin{equation}
M\equiv\frac{1}{\sqrt{16\pi G}}=\frac{M_{\rm Planck}}{\sqrt{16\pi}}\,,
\end{equation}
is related to the Planck mass $M_{\rm Planck}=1/\sqrt{G}=1.2\times10^{19}\,\textrm{GeV}$, the fundamental energy scale of gravitational interactions.
In this paper we will often refer to $M$ directly as the Planck mass.

\subsection{Bare action}

From an EFT point of view many terms can be added to the Einstein--Hilbert action, the first candidates being
the curvature squared terms \cite{Stelle:1977ry,Simon:1990ic}.
The most general four dimensional local 
action consistent with diffeomorphism invariance can be written as \cite{Donoghue:1994dn,Burgess:2003jk}
\begin{eqnarray}
S_{\rm eff}[g]  &=&  \int d^{4}x\sqrt{g}\Big[M^{4}\underbrace{c_{0}}_{\partial^{0}}+M^{2}\underbrace{c_1 R}_{\partial^2}+\underbrace{c_{2,1}R^{2}+c_{2,2}\textrm{Ric}^{2}+c_{2,3}\textrm{Riem}^{2}+c_{2,4} \square R}_{\partial^{4}}
\nonumber\\
&&\qquad+\frac{1}{M^{2}}\underbrace{\Big(c_{3,1}R\square R+c_{3,2}R_{\mu\nu}\square R^{\mu\nu}+c_{3,3}R^{3}+...\Big)}_{\partial^{6}}\Big] + O\!\left(\frac{\partial^8}{M^4}\right)\,,
\label{AG}
\end{eqnarray}
where we have emphasized the number of derivatives present in the various curvature invariants.
We have also defined the short hands $\textrm{Riem}^{2} \equiv R_{\mu\nu\alpha\beta}R^{\mu\nu\alpha\beta}$ and $\textrm{Ric}^{2} \equiv R_{\mu\nu}R^{\mu\nu}$.
The first two orders $\partial^0$ and $\partial^2$ correspond to a definite number of curvatures, explicitly $\mathcal{R}^0$ and $\mathcal{R}^1$.
At $O(\partial^4)$, the two counts start to mix such that at a given order in the derivatives, we find different orders in the curvatures, for instance, the first three terms are $\mathcal{R}^2$ terms while the last one is $\mathcal{R}^1$ term. 
Later we will see that the local heat kernel expansion is an expansion in the derivatives, while the non--local heat kernel expansion is an expansion in the curvatures.   

Basically all dimensionfull couplings in the theory are written in Planck units where the pure numbers $c_{i}$ are phenomenological bare parameters.
These parameters can be interpreted as the values of the dimensionless couplings at the Planck scale. 
Since their values are not known, we will later eliminate them in favour of the renormalized couplings which have to be measured in some experiment or observation at { some characteristic scale much smaller than the Planck scale.}
%
The constants entering the Einstein--Hilbert action are obviously
\begin{equation}
c_0 M^2 = 2\Lambda \qquad\qquad c_1 M^2 = - \frac{1}{16\pi G}\,,
\label{LG}
\end{equation}
where $\Lambda$ is the bare cosmological constant while $G$ is the bare Newton's constant.
Since we are performing an expansion in the Planck mass we have to normalize $c_{1}\equiv-1$.
The coefficient $c_{0}$ is somehow special from the EFT point of view since it comes to the left 
of the leading two derivatives interaction and is indeed enhanced by a power of $M^2$ rather than 
being suppressed by powers of $1/M^2$. This does not pose a problem as its renormalized value is either zero or 
exceedingly small, as inferred from the present cosmological observations.
Thus $c_{0}M^2 \ll 1$ and therefore, it can be considered as a mass term to the leading $\partial^2$ interactions.

We will group all the terms of the same order in derivatives together and rewrite (\ref{AG}) as
\begin{equation}
S_{\rm eff}[g] = M^{2}\left[I_{EH}[g]+\frac{1}{M^{2}}I_{2}[g]+\frac{1}{M^{4}}I_{3}[g]+...\right]\,,
\label{AG2}
\end{equation}
where we have defined
\begin{equation}
I_{EH}[g] \equiv M^2 I_0[g]+I_1[g]\,,\label{EH}
\end{equation}
to denote the classical Einstein--Hilbert action.
Note that now the action, apart the overall $M^2$ factor that we will use for the saddle point expansion, is written as a sum of terms of the form $\partial^{2(n+1)}/M^{2n}$ that makes explicit the energy/Planck mass expansion underlying the EFT approach.

In $d=4$, it is convenient, and physically meaningful, to rewrite the curvature square invariants by introducing the Euler density $E=\textrm{Riem}^{2}-4\textrm{Ric}^{2}+R^{2}$ and the Weyl tensor, the square of which is $C^{2}=E+2\textrm{Ric}^{2}-\frac{2}{3}R^{2}$.
The following relations allow us to switch between the Riemann basis $\{\textrm{Riem}^{2}, \textrm{Ric}^{2}, R^2\}$ and the Weyl basis $\{C^{2}, R^2, E\}$,
\begin{eqnarray}
\textrm{Riem}^{2}=-E+2C^{2}+\frac{1}{3}R^{2}\qquad\qquad\textrm{Ric}^{2}=\frac{1}{2}C^{2}-\frac{1}{2}E+\frac{1}{3}R^{2}\,.
\label{R2_1}
\end{eqnarray}
We can now rewrite $I_{2}$ in the Weyl basis
\begin{eqnarray}
I_{2}[g]
& = & \left(c_{2,1}+\frac{1}{3}c_{2,2}+\frac{1}{3}c_{2,3}\right)R^{2}+\left(\frac{1}{2}c_{2,2}+2c_{2,3}\right)C^{2}-\left(\frac{1}{2}c_{2,2}+c_{2,3}\right)E+c_{2,4} \square R\nonumber \\
& \equiv & c_{R^{2}}R^{2}+c_{C^{2}}C^{2}+c_{E}E+c_{\square R} \square R\,.\label{AG_2}
\end{eqnarray}
Using the conventions of higher derivative gravity, we find $c_{R^{2}}\equiv \frac{1}{\xi}$, $c_{C^{2}}\equiv \frac{1}{2\lambda}$ and $c_{E}\equiv - \frac{1}{\rho}$ \cite{Stelle:1977ry}.
Later on we will also use the Ricci basis $\{\textrm{Ric}^{2}, R^2, E\}$, where one uses the Euler density to eliminate the Riemann tensor in favour of the Ricci tensor and Ricci scalar via $\textrm{Riem}^{2}=4\textrm{Ric}^{2}-R^{2}+E$.

The six derivative term $I_3$ is composed of all the operators appearing in Table \ref{FLR3}. There are ten different invariants, the first two of them are of second order in the curvatures but contain a $\square$ operator that makes them of  $O(\partial^6)$. The remaining ones have three curvatures, in four dimensions the last two are not independent, so there is only one invariant with three Riemann tensors, or equivalently with three Weyl tensors. This is the Goroff--Sagnotti invariant that characterizes the perturbative two loops UV divergencies of quantum gravity \cite{Goroff:1985sz,Goroff:1985th,vandeVen:1991gw}. Note that, as we will also discuss later, when one includes matter, the suppression factor of $I_{3}$ in (\ref{AG2}) has to be replaced by $\frac{1}{M^{4}}\to\frac{1}{m^{2}M^{2}}$,
where $m$ is the mass of the lightest particle that has been integrated out.

\subsection{Effective action}

The covariant construction of the EFT focuses on the effective action $\Gamma[g]$, from which one then obtains the effective or quantum EOM
\begin{equation}
\frac{\delta\Gamma[g]}{\delta g_{\mu\nu}} = 0\,.
\label{EOM}
\end{equation}
The solution to this equation is the vacuum or background geometry, around which small quantum fluctuations are quantized. It is also the condition satisfied by the on--shell metric.
The virtue of the covariant formalism is that it allows for solutions of (\ref{EOM}) other than the flat Minkowski metric, such as static spherically symmetric or homogeneous isotropic metrics. In other words, to construct the EFT, we only assume that we are deep into the broken phase and the choice of the vacuum is not specified.  It indeed emerges as a solution to (\ref{EOM}) and the boundary conditions are ultimately provided by  experiments.
The effective action $\Gamma[g]$ is also the generating function of one--particle--irreducible (1PI) vertices. These are computed by taking functional derivatives with respect to the metric and/or matter fields and then going on--shell. As any scattering process or correlation function can be decomposed in 1PI parts, the knowledge of the effective action allows their determination on any background geometry.

As mentioned earlier, the basic underline assumption of EFT is that fluctuations are small which are then quantized in the standard way around an arbitrary background and this procedure is well under control even in the case of quantum gravity.  What we lack is a theory of large quantum spacetime fluctuations that are expected to be large starting from the scales of $O(M)$ which will allow transitions between different vacua of the theory. The best way to quantize the theory keeping the background unspecified is the background field method, which has been applied to gravity since the early times. We will work in Euclidean signature and explain later on how to perform the continuation to Lorentzian signature.  Within the background field method, the effective action is then defined by the following functional integral
%
\begin{figure}[!t]
\begin{center}
\includegraphics[scale=0.45]{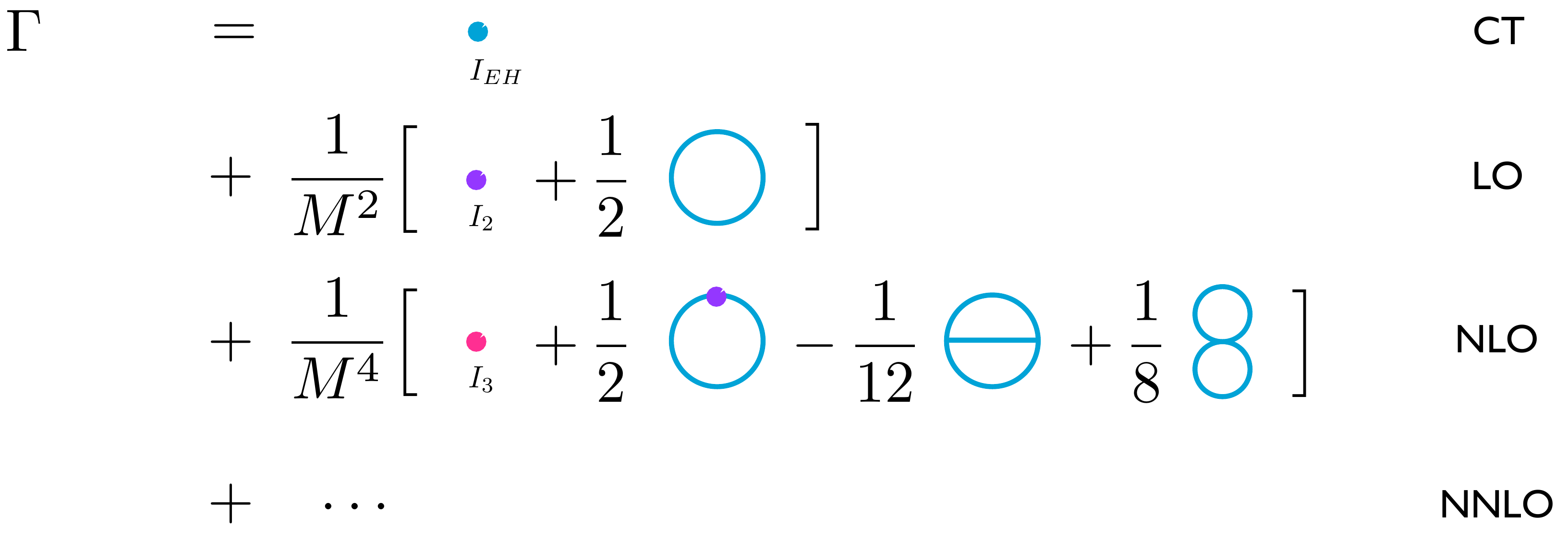}
\end{center}
\caption{Diagrammatic representation of the effective action of low energy quantum gravity, as in (\ref{EAG}). While the dots of different colors are self-explanatory, the blue line indicates the propagator with the Einstein--Hilbert term.}
\label{master}
\end{figure}
\begin{equation}
e^{-M^2 \Gamma[g]} = \int_{\rm 1PI} \mathcal{D} h_{\mu \nu} \,e^{-S_{\rm eff}[g+\frac{1}{M}h]}\,,
\label{bEA}
\end{equation}
where (\ref{exp_metric}) is implicitly used.
In (\ref{bEA}) the background gauge fixing and background ghost are understood, i.e. they can be seen as part of the definitions of the gauge invariant measure $\mathcal{D} h_{\mu \nu}$. As usual, the integral is only over 1PI diagrams. More details on the background field method in quantum gravity can be found in \cite{Codello:2008vh}.
In this setup the EFT effective action for gravity is computed via the saddle point, or loop expansion in the small parameter $1/M^2$.
We then substitute (\ref{AG2}) into (\ref{bEA}) to find 
\begin{equation}
e^{-M^2 \Gamma[g]} = \int_{\rm 1PI} \mathcal{D} h_{\mu \nu} \,e^{-M^2 \left\{ I_{EH}[g+\frac{1}{M}h] + \frac{1}{M^2} I_{2}[g+\frac{1}{M}h] + \cdots \right\}}\,.
\label{bEA2}
\end{equation}
Expanding now the invariants in the exponential and then the exponential, all in powers of $1/M$ gives a series of Gaussian integrals, with quadratic action $\int \sqrt{g}\,h \cdot I_{EH}^{(2)}[g] \cdot h$.
These are performed with the aid of Wick's theorem, leading to 
\begin{eqnarray}
\Gamma[g]  & = &  I_{EH}[g] + \frac{1}{M^{2}}I_{2}[g] + \frac{1}{M^{4}}I_{3}[g]+\cdots \nonumber \\
&& +\,\frac{1}{M^2}\frac{1}{2}\textrm{Tr}\log \left\{I_{EH}^{(2)}[g] + \frac{1}{M^{2}}I_{2}^{(2)}[g] + \frac{1}{M^{4}}I_{3}^{(2)}[g]+\cdots\right\}\nonumber \\
&& + \cdots\,,
\label{EAG0}
\end{eqnarray}
which can also be obtained by the standard loop expansion applied to the action $S_{\rm eff}[g]/M^2$.
We then expand the loops in $1/M^2$ and collect all terms of the same order to find
\renewcommand{\arraystretch}{2}
\begin{equation}
\begin{array}{cclc}
\Gamma[g] & = & I_{EH}[g] &{\bf CT}\\
&  & +\,\frac{1}{M^{2}}\left\{ I_{2}[g]+\frac{1}{2}\textrm{Tr}\log I_{EH}^{(2)}[g]\right\} &{\bf LO}\\
&  & +\,\frac{1}{M^{4}}\left\{ I_{3}[g]+\frac{1}{2}\textrm{Tr}\Big[\big(I_{EH}^{(2)}[g]\big)^{-1}I_{2}^{(2)}[g]\Big]+\textrm{2 loops with }I_{EH}[g]\right\} &{\bf NLO}\\
& & +\,O\!\left(\frac{1}{M^{6}}\right) &{\bf NNLO}\\
\end{array}
 \label{EAG}
\end{equation}
This is the final result for the covariant effective action within the EFT of gravity.
It is a covariant expression that can be used on all backgrounds which at classical level reproduces GR,
while the second line, the LO corrections, may contain all accessible quantum gravity phenomenology.
The evaluation of the LO corrections is the main goal of this paper.
A diagrammatic representation of this equation is given in Figure \ref{master}.
Before we proceed, a few comments about our main expression (\ref{EAG}) are in order:
\begin{itemize}
\item
It clearly shows why GR is so successful: the first quantum corrections are suppressed by Planck mass $1/M^{2}$.
\item
The fact that the propagator is given by the inverse of the Hessian of Einstein--Hilbert action guarantees unitarity.
\item
By construction, it is invariant under gauge transformations of the background field, the numerical coefficients which multiply the generally covariant quantities contributing to it  will nevertheless be different in different gauges, only on--shell $\Gamma[g]$, or the 1PI vertices derived from it, will be fully gauge invariant\footnote{In particular, also the on--shell metric can be gauge dependent. For a clear example of this see \cite{Dalvit:1997yc}.}.
\item
In the background field approach, the meaning of going on--shell depends on the order considered. At LO, the background metric is the solution of the classical Einstein EOM, at NLO the background metric is the solution of the LO EOM, and so on order by order\footnote{This is not the case if dealing with a perturbative or loop expansion of the partition function $Z$, or equivalently of the functional $W=\log Z$, where on--shell always refers to the tree level or classical background.}.
\item
From this covariant expansion, one can immediately recover the standard rules of EFT \cite{Donoghue:1992dd,Pich:1998xt,Rothstein:2003mp,Burgess:2007pt}. One finds that the general Lagrangian of order $E^2$ is to be used both at tree level and in loop diagrams; the general Lagrangian of order $E^{n \geq 4}$ is to be used at tree level and as an insertion in loop diagrams;
the renormalization program is carried out order by order.
\end{itemize}
Thus the expansion (\ref{EAG}) represents the basis for a well defined, consistent and predictive framework for low energy quantum gravity computations on an arbitrary background.
It is applicable when the relevant energy scales are much smaller than the Planck scale $E \ll M$, i.e. the scales which we can probe, or hope to probe, directly or indirectly in the near future, and where possible quantum gravitational phenomena may be hiding.
   
Finally, if we were to restore $\hbar$ in equation (\ref{EAG}), there will be a factor of $\hbar$ for each loop. For instance, the one loop diagrams in LO and NLO terms will be multiplied by $\hbar$, the two loop diagrams in NLO will be multiplied by $\hbar^2$.
But since both $I_2$ and $I_3$ are renormalized, their coefficients are also of order $\hbar$ and $\hbar^2$, respectively. More precisely, the bare couplings are of quantum origin since they encode the information about quantum gravity in the UV. From this point of view, the LO corrections are of order $\left(\frac{\hbar}{M^2}\right)$ while the NLO corrections are of order $\left(\frac{\hbar}{M^2}\right)^2$ and the EFT expansion is truly an expansion in inverse powers of the Planck mass rather than in powers of Newton's constant $G$\footnote{For higher loops, the bare couplings are expanded similarly in powers of $\frac{\hbar}{M^2}$.}.

\subsection{What we know}

We can use equation (\ref{EAG}) to reorganize many quantum gravity results by asking the question:
what do we  already know about the LO and NLO corrections?

To start with,  the local terms $I_{EH}, I_2,...$ encode the double information regarding the values of the
phenomenological constants and their renormalization.
Since these terms are local, their renormalized coefficients have to be measured by an experiment or observation,
but the loops in each of the lines of (\ref{EAG}) are UV divergent, so
we first need to renormalize the theory, by absorbing these divergencies in the bare parameters.
The freedom to choose the renormalization scale then leads directly to the RG running of the couplings of these invariants.
The renormalizations of the operators in $I_{EH}$ and $I_{2}$ steaming from the LO loops have been studied since the works of \cite{'tHooft:1974bx}
and lead to a first indication that quantum gravity was not {\it perturbatively} renomalizable.
From the EFT point of view, these are instead positive results, since they tell us how the gravitational couplings corresponding to the lowest operators renormalize.   
In a later section we will study in detail the renomalization of these operators.

The first finite part of the LO terms are what we can call the ``leading logs". They are directly
related to the UV divergencies of the $I_2$ operators. 
These non--local corrections, of the form $R \log \frac{-\square}{\mu^2} R$, are known since long \cite{Barvinsky:1987uw,Barvinsky:1990up,Dalvit:1994gf,Elizalde:1995tx,Gorbar:2002pw,Gorbar:2003yt,Espriu:2005qn,Satz:2010uu,Codello:2011js,Calmet:2015dpa} and have been discussed recently in the context of EFT in \cite{Donoghue:2014yha,Maggiore:2015rma}.
But also leading non--analytical $R^2 \log R$ terms may be present  \cite{Avramidi:2000bm,Alavirad:2013paa,Codello:2014sua,Ben-Dayan:2014isa,Rinaldi:2014gha}.
We will review these terms and their derivation in section \ref{section_nlna}.
A curvature expansion of the LO corrections can be performed systematically by employing the non--local heat kernel expansion, as pioneered in \cite{Barvinsky:1987uw,Barvinsky:1990up,Barvinsky:1990uq,Avramidi:2000bm,Dalvit:1994gf}.
In section \ref{section_curvature} we will evaluate all curvature squared ($\mathcal{R}^{2}$) terms  present in the LO corrections. 
In principle also the $\mathcal{R}^{3}$ terms are computable using the results presented in \cite{Barvinsky:1993en,Barvinsky:1994hw}, but we will not discuss them here.

Some of the LO terms can be obtained by integrating the conformal anomaly,
but other than in two dimensions, this does not lead to the full knowledge of the effective action.
The Reigert action \cite{Riegert:1984kt} captures these contributions, but in $d=4$ there is always an ambiguity in defining the conformal anomaly action and other options are available \cite{Barvinsky:1995it}.
From the perspective of the curvature expansion the anomaly induced terms start to be present at order $\mathcal{R}^3$ and higher, even if matter is present \cite{Donoghue:2015xla}.
In this paper we will focus on the $\mathcal{R}^2$ terms, so we will leave to a further work their systematic incorporation in the EFT framework. 
The phenomenology of these terms has been studied in great detail (for a review see \cite{Shapiro:2008sf,Mottola:2010gp} and reference therein) but we stress here that in general these terms are not the only ones in the effective action, since those not induced by the conformal anomaly will also be present simultaneously.

The Minkowski space EFT of gravity gives us indirect information about (\ref{EAG}).
The results of these computations are in principle obtainable from the vertices of the effective action evaluated on flat space. 
If we know the effective action to order $\mathcal{R}^{4}$, by taking four functional derivatives
with respect to the metric and setting $g_{\mu\nu} = \eta_{\mu\nu}$, we will be able to reproduce
the four graviton amplitude originally obtained in \cite{Dunbar:1994bn}.
Conversely, one can imagine to covariantize these results to infer the form of the LO
$\mathcal{R}^{3}$ or $\mathcal{R}^{4}$ terms, as done for the  $\mathcal{R}^{2}$ terms in \cite{Donoghue:2014yha}.

All this was about the LO terms, what do we know about the NLO ones?
Here the only known result is the famous computation of the two loop divergencies induced by $I_{EH}$  \cite{Goroff:1985sz,Goroff:1985th,vandeVen:1991gw}
that dictates the renormalization of the couplings in $I_3$,
while nothing is known about the finite part of these NLO corrections.

\subsection{Explicit form of LO corrections}

Before introducing matter in the next section, we take a moment to discuss explicitly
the trace--log formula that characterizes the LO corrections in (\ref{EAG}) for pure gravity, that from now on we will call $T_2\equiv \frac{1}{2} \textrm{Tr} \log I_{EH}^{(2)}[g]$.
The Hessian of the Einstein--Hilbert action is a differential operator and for a proper choice of the gauge fixing it is of Laplace type.
This has very interesting consequences, as it allows the use of heat kernel techniques, both local and non--local.
Although, at present, we do not have the mathematical technology to fully compute the trace--log on an arbitrary background,
exact results are available only on some specific spacetimes, like maximally symmetric spaces \cite{Avramidi:2000bm}.

Skipping the details of the computations, which can be found elsewhere (see for example \cite{Codello:2008vh}),
one finds
\begin{equation}
T_2=\frac{1}{2} \textrm{Tr} \log (\Delta_{2}-2\Lambda)-\textrm{Tr} \log \Delta_{gh}\,,
\label{PGH_1}
\end{equation}
where the differential operator for the spin two part is
\begin{equation}
(\Delta_{2})^{\mu\nu}_{\alpha\beta}=-\square \delta_{\alpha\beta}^{\mu\nu}+W^{\mu\nu}_{\alpha\beta}\,,
\label{laplacian2}
\end{equation}
where $\square=g^{\mu\nu}\nabla_{\mu}\nabla_{\nu}$ is the covariant
Laplacian, $\delta_{\rho\sigma}^{\mu\nu}=\frac{1}{2}\left(\delta_{\rho}^{\mu}\delta_{\sigma}^{\nu}+\delta_{\sigma}^{\mu}\delta_{\rho}^{\nu}\right)$
is the symmetric spin two tensor identity and
where we have defined the tensor
\begin{eqnarray}
W_{\rho\sigma}^{\alpha\beta} & = & \left(\delta_{\rho\sigma}^{\alpha\beta}-\frac{1}{2}g^{\alpha\beta}g_{\rho\sigma}\right)R+g^{\alpha\beta}R_{\rho\sigma}+R^{\alpha\beta}g_{\rho\sigma}\nonumber \\
 &  & -\frac{1}{2}\left(\delta_{\rho}^{\alpha}R_{\sigma}^{\beta}+\delta_{\sigma}^{\alpha}R_{\rho}^{\beta}+R_{\rho}^{\alpha}\delta_{\sigma}^{\beta}+R_{\sigma}^{\alpha}\delta_{\rho}^{\beta}\right)-\left(R_{\;\;\rho\;\,\;\sigma}^{\alpha\;\;\beta}+R_{\;\;\sigma\;\,\;\rho}^{\alpha\;\;\beta}\right)\nonumber \\
 &  & -\frac{d-4}{2(d-2)}\left(-R\, g^{\alpha\beta}g_{\rho\sigma}+2g^{\alpha\beta}R_{\rho\sigma}\right)\,.
\label{PGH_2}
\end{eqnarray}
The effective ``gravitational'' mass is { $m^{2}\equiv-2\Lambda$.}
When we add matter (in terms of a scalar field $\phi$) there will be an extra term proportional to the effective potential $V(\phi)$ evaluated at the minimum $v$ divided by $M^2$, i.e. {  $m^2=-2\Lambda +V(v)/M^2$.
Here and in the following, in absence of a cosmological constant and/or of an effective potential, we add an IR regularization mass $\mu^2$ as in the case of massless fields.}
%
For the spin one ghosts we instead have the following differential operator
\begin{equation}
(\Delta_{gh})_{\nu}^{\mu}=-\square\delta_{\nu}^{\mu}-R_{\nu}^{\mu}\,.
\label{PGH_4}
\end{equation}
We remark here that these relations are valid within the de Donder (harmonic) background gauge \cite{Codello:2008vh}.
The commutators of covariant derivatives $\Omega_{\mu\nu}\equiv[\nabla_{\mu},\nabla_{\nu}]$ are
\begin{eqnarray}
(\Omega_{\mu\nu}^{(2)})_{\alpha\beta}^{\rho\sigma}&=&-\frac{1}{2}\left(\delta^\rho_\alpha R_{\mu\nu\;\;\beta}^{\;\;\;\;\sigma}+\delta^\sigma_\alpha  R_{\mu\nu\;\;\beta}^{\;\;\;\;\rho} +\delta^\rho_\beta R_{\mu\nu\;\;\alpha}^{\;\;\;\;\sigma}+\delta^\sigma_\beta R_{\mu\nu\;\;\alpha}^{\;\;\;\;\rho} \right)\nonumber\\
\qquad\qquad(\Omega_{\mu\nu}^{(gh)})^{\beta}_\alpha&=& R_{\mu\nu\;\;\alpha}^{\;\;\;\;\beta}\,.
\end{eqnarray}
The specification of the two differential operators (\ref{laplacian2}) and (\ref{PGH_4})
completely determines the form of the finite part of the functional trace in (\ref{PGH_1})
and thus the form of the LO quantum corrections in the EFT of pure gravity.
Unfortunately, as mentioned earlier, we don't know how to exactly compute these traces on an arbitrary background
and one has to resort to approximations as those provided by the heat kernel expansion.
 
\section{EFT of gravity coupled to matter}\label{section_matter}

In physical cases matter is always present and therefore the EFT must also include the relative field fluctuations.
This leads to, for example, the terms in the effective action responsible for the quantum corrections to Newton's potential.

\subsection{Bare action}

For simplicity and illustrative purpose we only consider the case of a real scalar. 
The most general form of matter invariants with at most two derivatives can be written as
\begin{equation}
I_{S}[\phi,g]  =  \int d^{4}x\sqrt{g}\left[\frac{1}{2}Z(\phi)g^{\mu\nu}\partial_{\mu}\phi\partial_{\nu}\phi+V(\phi)+F(\phi)R \right]\,,
\label{IS}
\end{equation}
where we impose $V(0)=F(0)=0$. We can use a field redefinition to eliminate $Z(\phi)$ which is set to unity in the following.
The general form for the bare action when matter is present can be written as
\begin{equation}
S_{\rm eff}[\phi,g]  = M^{2}\left[I_{EH}[g]+\frac{1}{M^{2}}I_{S}[\phi,g]+\frac{1}{M^{2}}I_{2}[g]+\frac{1}{M^{4}}I_{S,4}[\phi,g]+\frac{1}{M^{4}}I_{3}[g]+...\right]\,,
\label{AMG}
\end{equation}
where we have included the four derivative terms $I_{S,4}[\phi,g]$ which will be generated by the loops.

\subsection{Effective action}

With respect to the pure gravity case, the evaluation of the effective action needs a couple of steps more since we need to deal with the field multiplet traces.
In particular, we need to expand the multiplet trace in the trace--log formula in powers of the inverse Planck  mass\footnote{We are expanding as $(g_{\mu\nu},\phi)\rightarrow(g_{\mu\nu},\phi)+\frac{1}{M}(h_{\mu\nu},M\chi)$.}
\begin{eqnarray*}
T & = & \frac{1}{2}\textrm{Tr}\log\left(\begin{array}{cc}
I_{EH}^{hh}+\frac{1}{M^2}I_{S}^{hh} & \frac{1}{M} I_{S}^{h\chi}\\
\frac{1}{M} I_{S}^{\chi h} & I_{S}^{\chi\chi}
\end{array}\right)\\
 & = & \frac{1}{2}\textrm{Tr}\log\left[\left(\begin{array}{cc}
I_{EH}^{hh}+\frac{1}{M^2}I_{S}^{hh} & 0\\
0 & I_{S}^{\chi\chi}
\end{array}\right)+\frac{1}{M}\left(\begin{array}{cc}
0 & I_{S}^{h\chi}\\
I_{S}^{\chi h} & 0
\end{array}\right)\right]\\
& \equiv &\frac{1}{2} \textrm{Tr}\log\left[A+\frac{1}{M} B\right]\,,
\end{eqnarray*}
using the expansion $\log\left[A+\frac{1}{M}B\right]
 = \log A+\frac{1}{M}\, A^{-1}B-\frac{1}{2M^2}(A^{-1}B)^2+O\left(\frac{1}{M^3}\right)$.
From the relation
\begin{eqnarray*}
A^{-1}B =  
\left(\begin{array}{cc}
\frac{1}{I_{EH}^{hh}+\frac{1}{M^2}I_{S}^{hh}} & 0\\
0 & \frac{1}{I_{S}^{\chi\chi}}
\end{array}\right)\left(\begin{array}{cc}
0 & I_{S}^{h\chi}\\
I_{S}^{\chi h} & 0
\end{array}\right)
 =
 \left(\begin{array}{cc}
0 & \frac{1}{I_{EH}^{hh}+\frac{1}{M^2}I_{S}^{hh}}I_{S}^{h\chi}\\
\frac{1}{I_{S}^{\chi\chi}}I_{S}^{\chi h} & 0
\end{array}\right)
 \,,
\end{eqnarray*}
we see that the linear term traces to zero, i.e. $\textrm{tr}(A^{-1}B) = 0$,
while the quadratic one gives
\begin{eqnarray*}
\textrm{tr}(A^{-1}B)^2 
 & = & \frac{1}{I_{EH}^{hh}+\frac{1}{M^2}I_{S}^{hh}}I_{S}^{h\chi}\frac{1}{I_{S}^{\chi\chi}}I_{S}^{\chi h}+\frac{1}{I_{S}^{\chi\chi}}I_{S}^{\chi h}\frac{1}{I_{EH}^{hh}+\frac{1}{M^2}I_{S}^{hh}}I_{S}^{h\chi}\,.
\end{eqnarray*}
Thus, after the multiplet trace, the trace--log becomes
\begin{eqnarray}
T & = & \frac{1}{2}\textrm{Tr}\log\left[I_{EH}^{hh}+\frac{1}{M^2}I_{S}^{hh}\right]+\frac{1}{2}\textrm{Tr}\log I_{S}^{\chi\chi}\nonumber\\
&&\qquad\qquad-\frac{1}{2M^2}\textrm{Tr}\left[\frac{1}{I_{EH}^{hh}+\frac{1}{M^2}I_{S}^{hh}}I_{S}^{h\chi}\frac{1}{I_{S}^{\chi\chi}}I_{S}^{\chi h}\right]
+O\!\left(\frac{1}{M^{4}}\right)\,,
\label{multitrlog}
\end{eqnarray}
where we used the fact that $\textrm{tr}(A^{-1}B)^3=0$.
Upon expanding the remaining $1/M^2$ in the functional traces we finally find the form of the effective action for a scalar--gravity EFT
\renewcommand{\arraystretch}{2}
\begin{equation}
\begin{array}{cclc}
\Gamma & = & I_{EH} &{\bf CT}\\
&  & +\,\frac{1}{M^{2}}\left\{ I_{2}+I_{S} +\frac{1}{2}\textrm{Tr}\log I_{EH}^{hh}+\frac{1}{2}\textrm{Tr}\log I_{S}^{\chi\chi}\right\} &{\bf LO}\\
&  & +\,\frac{1}{M^{4}}\left\{I_{3}+I_{S,4}+\frac{1}{2}\textrm{Tr}\frac{1}{I_{EH}^{hh}}I_{S}^{hh}-\frac{1}{2}\textrm{Tr}\frac{1}{I_{EH}^{hh}}I_{S}^{h\chi}\frac{1}{I_{S}^{\chi\chi}}I_{S}^{\chi h} + \textrm{2 loops with }I_{EH}\right\} &{\bf NLO}\\
& & +\,O\!\left(\frac{1}{M^{6}}\right)&{\bf NNLO}\\
\end{array}
 \label{EAMG}
\end{equation}
\begin{figure}[!t]
\begin{center}
\includegraphics[scale=0.45]{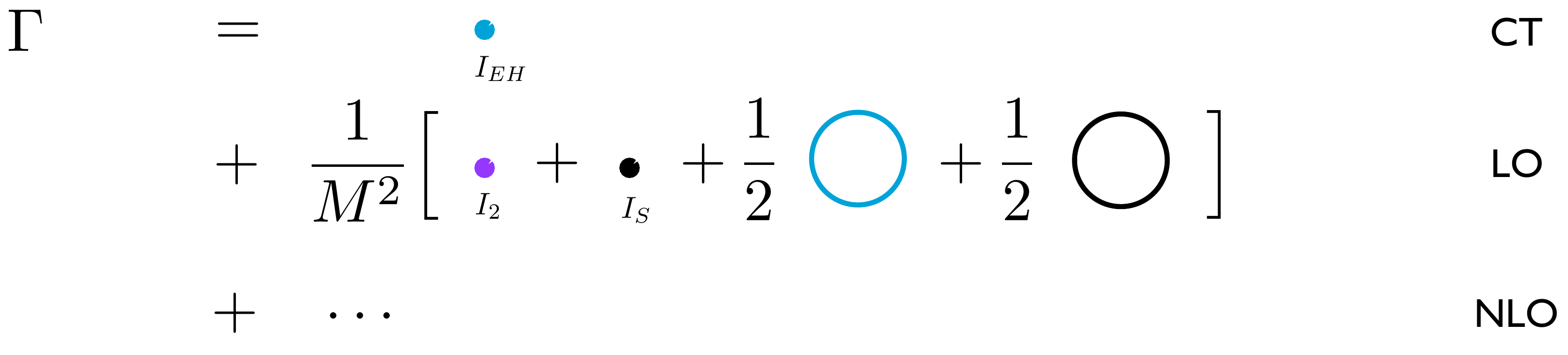}
\end{center}
\caption{Diagrammatic representation of the low energy EFT expansion of the effective action when matter, in the form of a real scalar field, is present.}
\label{mastermatter}
\end{figure}
%
\!\!\!Note that $I_{S}$ is ``Planck suppressed", as expected by the fact that in GR the energy--momentum tensor of matter is multiplied by $1/M^2$ in Einstein's equations.
Thus classical matter appears at the same order as the LO quantum corrections (apart from a factor of $\hbar$). A diagrammatic representation of this equation is given in Figure \ref{mastermatter}.

The scalar--gravity NNLO terms, not shown in equation (\ref{EAMG}), will contain the second order expansion in $1/M^2$ of the first trace--log in equation (\ref{multitrlog}) and the first order term of the last trace of the same equation together with the contribution from $\textrm{tr}(A^{-1}B)^4$. All these terms contribute to the corrections of Newton's potential along with the LO and NLO terms \cite{Donoghue:1993eb}. In fact, after restoring $\hbar$, the corrections to Newton's potential are of order $\hbar/M^2$ and, when compared to the LO corrections of order $\hbar$, are suppressed by an additional factor of $1/M^2$.
Thus there exist possible quantum gravitational effects which are less suppressed than the known corrections to Newton's potential, but these cannot be observed on Minkowski space and are indeed present on a spacetime with non--zero curvature, as for example FRW or Schwarzschild. This peculiar observation is one main reason that motivates the analysis of the LO corrections in equation (\ref{EAMG}) and their possible physical implications.

\subsection{What we know}

As in the case of pure gravity, we can now summarize what is already known about equation (\ref{EAMG}). 
UV divergencies have been computed for many matter gravity combinations \cite{'tHooft:1974bx,Deser:1974cz,Deser:1974cy,Deser:1974zzd}. Although it was shown originally  that, despite one loop gravity was finite on--shell, as soon as, matter was introduced this was not the case anymore.
As in the pure gravity case, these results describe the RG flow of the relative couplings.

As said earlier, the computation of the corrections to Newton's potential have been performed on Minkowski space using Feynman diagrams techniques \cite{Donoghue:1993eb,Donoghue:1994dn,Akhundov:1996jd,Khriplovich:2002bt,BjerrumBohr:2002kt} and further verified by modern methods \cite{Bjerrum-Bohr:2013bxa}. These corrections can also be derived by the 1PI diagrams obtained by the effective action (\ref{EAMG}) if all the relevant terms upto NNLO are considered.   
The polarization diagrams involved are covariantly described by the leading logs introduced in the pure gravity case, while the new non trivial part of the computation involves the matter--graviton vertex and four matter vertex in the presence of gravitons. The part of the leading logs of the matter--graviton vertex have now been computed on an arbitrary background \cite{Codello:2015oqa}. The full computation of all the relevant 1PI diagrams is left for future work.

\subsection{Minimally and conformally coupled matter}

Equation (\ref{EAMG}) shows that matter induced LO corrections are approximated by the trace--log formula. When matter is evaluated in its vacuum configuration, it becomes as minimally coupled and/or conformally coupled, that we consider in the following without loss of generality.
The bare action for a real scalar field, a Dirac spinor and an abelian gauge field is given by 
\begin{equation}
I_{m}[\phi,\psi,A_{\mu},\bar{c},c,g]=I_0[\phi,g]+I_\frac{1}{2}[\psi,g]+I_1[A_{\mu},\bar{c},c,g]\,,
\label{mmm_1}
\end{equation}
where
\begin{eqnarray}
I_0[\phi,g]&=&\int d^{4}x\sqrt{g}\left\{ \frac{1}{2}\partial_{\mu}\phi\partial^{\mu}\phi+\frac{\chi}{12}\phi^{2}R+\frac{m_{\phi}^{2}}{2}\phi^{2}\right\}\nonumber\\
I_\frac{1}{2}[\psi,g]&=&\int d^{4}x\sqrt{g}\,\bar{\psi}(\slashed{\nabla}+m_{\psi})\psi\nonumber\\
I_1[A_{\mu},\bar{c},c,g]&=&\int d^{4}x\sqrt{g}\left\{\frac{1}{4}F_{\mu\nu}F^{\mu\nu}+\frac{1}{2\alpha}\left(\partial_{\mu}A^{\mu}\right)^{2}+\partial_{\mu}\bar{c}\,\partial^{\mu}c\right\} \,.
\end{eqnarray}
Here $\chi$ is a parameter, only when $\chi=1$ the scalar action is conformal invariant (if the scalar
has conformal weight one). The Dirac operator is defined using the covariant Dirac matrices $\gamma^{\mu}=e_{a}^{\mu}\gamma^{a}$ and the vierbein formalism.
Note that on an arbitrary curved manifold the abelian ghosts do not
decouple and cannot be discarded. In what follows, we will choose the gauge $\alpha=1$.
The computation of the Hessian is straightforward and gives the following matter traces
%
\begin{eqnarray}
T_0&=&\frac{1}{2} \textrm{Tr} \log (\Delta_{0}+m^2_\phi) \nonumber \\
T_\frac{1}{2}&=&-\frac{1}{2} \textrm{Tr} \log (\Delta_\frac{1}{2}+m^2_\psi) \nonumber \\
T_1&=&\frac{1}{2} \textrm{Tr} \log \Delta_{1}-\textrm{Tr} \log (-\square)\,,
\label{Tm}
\end{eqnarray}
where the Laplacians are
\begin{equation}
\Delta_{0}=-\square+\frac{\chi}{6}R\qquad\qquad\Delta_{\frac{1}{2}}=-\square+\frac{1}{4}R\qquad\qquad
(\Delta_{1})^{\mu}_{\nu}=-\square \delta^{\mu}_{\nu}+R^{\mu}_{\nu}
\,.\label{mmm_3}
\end{equation}
The commutators of the covariant derivatives $\Omega_{\mu\nu}\equiv[\nabla_{\mu},\nabla_{\nu}]$ are
\begin{equation}
\Omega_{\mu\nu}^{(0)}=0
\qquad\qquad\Omega_{\mu\nu}^{(\frac{1}{2})}=\frac{1}{4}\gamma^{\alpha}\gamma^{\beta}R_{\alpha\beta\mu\nu}
\qquad\qquad(\Omega_{\mu\nu}^{(1)})^{\alpha}_{\;\;\beta}= R^{\alpha}_{\;\;\beta\mu\nu} \,.
\end{equation}
These last two relations are all we need to employ the heat kernel methods to compute the traces.

\section{Local terms via heat kernel}\label{section_renor}

We are now ready to start evaluating the LO corrections appearing in (\ref{EAG}) or in (\ref{EAMG}).
The task amounts to evaluating the trace--log of the Einstein--Hilbert action (\ref{PGH_1}), or of minimally coupled matter (\ref{EAMG}).
First we will compute the local part of the trace--log formula.
This will lead us to discuss the issue of renormalization and to review
the well known UV divergencies of quantum gravity.
Here we will see this problem from the point of view of the EFT of gravity, dictating how the gravitational couplings run.

In both the gravitational and the matter cases we need to compute the trace--log of the form 
\begin{equation}
T=\frac{1}{2}\textrm{Tr}\log (\Delta + m^2)\,,
\label{mmm_2}
\end{equation}
where $\Delta= -\square\,{\bf 1} + {\bf U}$ is the Hessian of $I_{EH}$, of the ghosts action, or of any of the matter parts
contained in $I_{m}$.
In the case of massless fields, we regularize IR divergencies by adding
a mass term $m=\mu$.
The trace in (\ref{mmm_2}) can be expanded in powers of the curvature
by employing the local heat kernel expansion which we review in the appendix \ref{heat-kernel}.
It is useful to rewrite the trace (\ref{mmm_2}) as
\begin{equation}
T=-\frac{1}{2}\int_{0}^{\infty}\frac{ds}{s}\, K(s)\, e^{-s m^2}\,,
\label{mmm_4}
\end{equation}
so that we can link it directly to the trace of the heat kernel $K(s)=\textrm{Tr}\, e^{-s\Delta} $
of the Laplacian operator $\Delta$.
We can now employ the local heat kernel expansion for a second order operator
\begin{equation}
K(s)= \frac{1}{(4\pi)^{d/2}}\sum_{n=0}^{\infty}s^{n-d/2}B_{2n}(\Delta) \,,
\label{hk_1}
\end{equation}
where the $B_{2n}$ are the integrated heat kernel coefficients,
the first few are known and we have collected them in the appendix \ref{heat-kernel}.
With the aid of this expansion, the trace can now be split into a divergent and a (local) finite part $T = T_{\rm div} + T_{\rm finite}$, the first will be discussed in the next section, while the latter in section \ref{section_finite_local}.

\subsection{Regularization}\label{section_reg}

When the expansion (\ref{hk_1}) is inserted into (\ref{mmm_4}) we encounter UV divergencies at the lower extrema of the integral.
In four dimensions, the first three terms will be divergent, in particular they will have quartic, quadratic and logarithmic divergencies.
To regularize (\ref{mmm_4}) we will employ for comparison, both dimensional regularization and a cutoff regularization.

\paragraph{Dimensional regularization}
In this case, even if we are interested in four dimensions, we keep $d$ unspecified so that (\ref{mmm_4}) is finite for non--integer dimensions.
We find\footnote{We recall the basic integral $\int_{0}^{\infty} ds\,s^{t-1}e^{-sm^{2}}=\Gamma(t)/m^{2t}$.}
\begin{equation}
T =  -\frac{1}{2(4\pi)^{d/2}}\sum_{n=0}^{\infty}B_{2n}\int_{0}^{\infty}ds\, s^{n-d/2-1}e^{-sm^{2}}
  = - \frac{1}{2(4\pi)^{d/2}}\sum_{n=0}^{\infty}m^{d-2n} \Gamma\!\left(n-\tfrac{d}{2}\right) B_{2n}\,.
\label{dimensional regularization_1}
\end{equation}
The Gamma function $\Gamma(x)$ has poles when $x=0,-1,-2,-3,...$ so we have
UV divergencies for $n-\frac{d}{2}\leq0$ for integer $d$.
As we will see in a moment, these dimensional regularization poles in $x=0,-1,-2,...$ correspond to
logarithmic, quadratic, quartic, ... divergencies in presence of a cutoff.
Finite terms start from $n=\frac{d}{2}+1$. 
We now set $d=4-\epsilon$ in (\ref{dimensional regularization_1}) and expand the Gamma functions
for small $\epsilon$ to find the following UV divergent contributions
\begin{equation}
T_{\rm div} =  -\frac{1}{2(4\pi)^2}\left[m^{4}\left(\frac{1}{\epsilon}+\frac{3}{4}-\frac{\gamma}{2}\right)B_{0}
+m^{2}\left(-\frac{2}{\epsilon}-1+\gamma\right)B_{2}
+\left(\frac{2}{\epsilon}-\gamma\right)B_{4}\right]\,.
\label{dimensional regularizationdiv}
\end{equation}
In dimensional regularization quartic and quadratic divergencies vanish for a massless
theory $m \rightarrow 0$ and only logarithm divergencies are left.
For a massive theory we instead have all of them.
At this point it is not clear what happens to the $B_{6},B_{8},...$ terms in equation (\ref{dimensional regularization_1}) in a
massless theory since the $m\rightarrow 0$ limit is (IR) divergent.
To understand this limit one needs the full non--local heat kernel that we will introduce in section \ref{section_curvature}.
If we do not worry about this fact and dropping the finite renormalization constant we find the classical result of 'tHooft for the logarithmic divergencies
of a four dimensional theory
\begin{equation}
T_{\rm div} = -\frac{1}{\epsilon}\frac{1}{(4\pi)^{2}}B_{4}\,.
\label{dimensional regularization_3}
\end{equation}
A theory is perturbatively renomalizable at  one loop if the operators present in $B_4$ are also present in the bare action.
In the EFT framework this is instead not a requirement and eventual divergencies will be renormalized
by switching on the operators appearing in $B_4$ not already present in the bare action.

\paragraph{Cutoff regularization}

We can instead apply a hard cutoff\footnote{Note that the UV scale $\Lambda_{\rm UV}$ can be equated to $M$, consistently with the fact that we used the Planck scale to make the couplings dimensionless in the bare action, which indeed should be interpreted as the effective action at scale $M$.} $\Lambda_{\rm UV}$ to (\ref{mmm_4}) and work directly in $d=4$
\begin{equation}
T=
-\frac{1}{2(4\pi)^{2}}\sum_{n=0}^{\infty}B_{2n}\int_{1/\Lambda_{\rm UV}^{2}}^{\infty}ds\, s^{n-3}e^{-sm^{2}}\,.
\label{cutoff_1}
\end{equation}
Starting at $n=3$ the integral is finite in the limit $\Lambda_{\rm UV} \to \infty$ and gives the same result as in the dimensional regularization case (\ref{dimensional regularization_1}).
Upon using the following integrals
\begin{eqnarray}
\int_{1/\Lambda_{\rm UV}^{2}}^{\infty}\frac{ds}{s^{3}}e^{-sm^{2}} & = & \frac{\Lambda_{\rm UV}^{4}}{2}-\Lambda_{\rm UV}^{2}m^{2}+\frac{m^{4}}{2}\left(\log\frac{\Lambda_{\rm UV}^{2}}{m^{2}}-\gamma+\frac{3}{2}\right)+O\!\left(\frac{1}{\Lambda_{\rm UV}^{2}}\right)\nonumber \\
\int_{1/\Lambda_{\rm UV}^{2}}^{\infty}\frac{ds}{s^{2}}e^{-sm^{2}} & = & \Lambda_{\rm UV}^{2}-m^{2}-m^{2}\left(\log\frac{\Lambda_{\rm UV}^{2}}{m^{2}}-\gamma\right)+O\!\left(\frac{1}{\Lambda_{\rm UV}^{2}}\right)\nonumber \\
\int_{1/\Lambda_{\rm UV}^{2}}^{\infty}\frac{ds}{s}e^{-sm^{2}} & = & \log\frac{\Lambda_{\rm UV}^{2}}{m^{2}}-\gamma+O\!\left(\frac{1}{\Lambda_{\rm UV}^{2}}\right)
\,,
\label{cutoff_2}
\end{eqnarray}
now gives the following UV divergent part
\begin{eqnarray}
T_{\rm div} & = & -\frac{1}{2(4\pi)^{2}}\left[\left(\frac{\Lambda_{\rm UV}^{4}}{2}-\Lambda_{\rm UV}^{2}m^{2}+\frac{m^{4}}{2}\log\frac{\Lambda_{\rm UV}^{2}}{m^{2}}+\frac{m^{4}}{2}\left(\frac{3}{2}-\gamma\right)\right)B_{0}\right.\nonumber \\
 &  & +\left(\Lambda_{\rm UV}^{2}-m^{2}(1-\gamma)-m^{2}\log\frac{\Lambda_{\rm UV}^{2}}{m^{2}}\right)B_{2}
 \left.+\left(\log\frac{\Lambda_{\rm UV}^{2}}{m^{2}}-\gamma\right)B_{4}\right]\,.\label{cutoff_4}
\end{eqnarray}
Thus the heat kernel coefficients $B_{0},B_{2}$ and $B_{4}$ are
the coefficients of quartic, quadratic and logarithmic
divergencies, respectively.
Again, to renormalize the theory we need the bare action to contain the invariants
present in $B_{0},B_{2}, B_{4}$. 

A comparison between the two regularization schemes shows that, to switch from cutoff to dimensional regularization, one needs to replace $\log \frac{ \Lambda_{\rm UV}}{m} \to \frac{1}{\epsilon}$  and set the other divergencies to zero $\Lambda_{\rm UV}^4 \to 0 $ and $ \Lambda_{\rm UV}^2  \to 0$.
Note that the finite renormalization constants are also the same in both schemes.
Clearly, the finite terms $B_{6}, B_{8}, ...$ are also independent of the regularization employed.

\subsection{Renormalization}

To renormalize the theory we absorb the UV divergencies into the bare couplings present in the actions $I_{EH}, I_2,...$ defining the renormalized couplings.
LO renormalizes CT and LO, while NLO renormalizes NLO, LO and CT and so on.
This will make the EFT finite at the cost of introducing a measured value for each divergent coupling,
but it will also tell us how the renormalized couplings depend on the arbitrary renormalization scale. 

In the appendix \ref{heat-kernel} we evaluate the heat kernel coefficients $B_4$ for all operators
under consideration. 
Using (\ref{dimensional regularization_3}) we find, for the gravitational trace, the following UV divergent part
in the Weyl basis
\begin{equation}
\left.T_{2}\right|_{\rm div}
 =  -\frac{1}{\epsilon}\frac{1}{(4\pi)^{2}}\int d^{4}x\sqrt{g}\left[\frac{7}{20}C^{2}+\frac{1}{4}R^{2}+\frac{149}{180}E-\frac{19}{15}\square R\right]\,. \label{Ren_1.1}
\end{equation}
In the Ricci basis the UV divergencies are instead
\begin{equation}
\left.T_2\right|_{\rm div} =  -\frac{1}{\epsilon}\frac{1}{(4\pi)^{2}}\int d^{4}x\sqrt{g}\left[\frac{7}{10}R_{\mu\nu}R^{\mu\nu}+\frac{1}{60}R^{2}+\frac{53}{45}E-\frac{19}{15}\square R\right]\,,
\end{equation}
which is the original result of \cite{'tHooft:1974bx,Hawking:1976ja}.
When using a cutoff regularization, or generally in the presence of masses, we also need $B_{0}$ and $B_{2}$, which we compute in appendix \ref{heat-kernel}.
The final result for the divergent part (modulo finite renormalization terms) is
\begin{eqnarray}
\left.T_{2}\right|_{\rm div} & = & -\frac{1}{2(4\pi)^{2}}\int d^{4}x\sqrt{g}\left[\Lambda_{\rm UV}^{4}-10\Lambda_{\rm UV}^{2}m^{2}+5m^{4}\log\frac{\Lambda_{\rm UV}^{2}}{m^{2}}\right.\nonumber \\
 &  & +\left(-\frac{23}{3}\Lambda_{\rm UV}^{2}+\frac{13}{3}m^{2}\log\frac{\Lambda_{\rm UV}^{2}}{m^{2}}\right)R\nonumber \\
 &  & \left.+\left(\frac{7}{20}C^{2}+\frac{1}{4}R^{2}+\frac{149}{180}E-\frac{19}{15}\square R\right)\log\frac{\Lambda_{\rm UV}^{2}}{m^{2}}\right]\,,
 \label{Ren_3}
\end{eqnarray}
where {$m^{2}=-2\Lambda+V(v)/M^2$.} Note that, both the graviton and the ghost contribute to the terms present  in the massless limit while the ghost does not contribute to the massive terms. 
We also dropped the finite renormalization constants. 
As mentioned earlier, to switch to dimensional regularization just replace $\log \frac{\Lambda_{\rm UV}}{m} \to \frac{1}{\epsilon}$ and set the other divergencies to zero.
We remark that these off shell coefficients are still gauge dependent \cite{Kallosh:1978wt,Capper:1983fu}.

We can now proceed to the explicit renormalization of the $I_{EH}$ and $I_{2}$ couplings.
From (\ref{Ren_3}), the explicit renormalization of the $I_{2}$ couplings, in the massless and dimensional regularization case is
\begin{equation}
c_{i}^{R}=c_{i}^{B}-\frac{1}{\epsilon}\frac{\gamma_{i}}{(4\pi)^{2}} \,,\label{Ren_5}
\end{equation}
where the coefficients $\gamma_{i}$ are reported in Table \ref{RenR2}.
This leads to the beta functions\footnote{The one--loop beta function coefficient is {\it minus} the coefficient of $\frac{1}{\epsilon}$.}
\begin{equation}
\mu \partial_\mu c_{i}(\mu)=\frac{\gamma_{i}}{(4\pi)^{2}}\,,
\label{BetaR2}
\end{equation}
which can be integrated to give
\begin{equation}
 c_{i}(\mu_2)=c_{i}(\mu_1)+\frac{\gamma_{i}}{(4\pi)^{2}} \log \frac{\mu_2}{\mu_1}\,,
\label{runR2}
\end{equation}
which relates the renormalized couplings at different renormalization scales\footnote{{It is clearly understood that the RG scale first used in equation (\ref{BetaR2}) and the IR regulator introduced in the previous sections to treat massless fields are different scales, even though we will use the same notation $\mu$ for both in order to keep the notation as simple as possible. }}.
This scale dependence of the phenomenological parameters is in principle an observable effect.

\renewcommand{\arraystretch}{2}
\begin{table}[t]
\begin{centering}
\begin{tabular}{cccca}
\hline\hline \\[-1.15cm]
 &\mc{1}{0} & \mc{1}{$\frac{1}{2}$} & \mc{1}{1} & \mc{1}{2} \tabularnewline
\hline
$\gamma_{C^{2}}$        & $\frac{1}{120}$ & $\frac{1}{20}$ & $\frac{1}{10}$ & $\frac{7}{20}$  \tabularnewline\hline
$\gamma_{R^{2}}$        & $\frac{(1-\chi)^{2}}{72}$ & $0$ & $0$ & $\frac{1}{4}$                       \tabularnewline\hline
$\gamma_{E}$              & $-\frac{1}{360}$ & $-\frac{11}{360}$ & $-\frac{31}{180}$ & $\frac{149}{180}$  \tabularnewline\hline
$\gamma_{\square R}$ & $-\frac{1}{30}$ & $-\frac{1}{30}$ & $\frac{1}{10}$ & $-\frac{19}{15}$  \tabularnewline
\hline\hline
\end{tabular}
\par\end{centering}
\caption{Renormalization constants for the four derivative operators in $I_{2}$ in the Weyl basis.
\label{RenR2}}
\end{table}

But also $\Lambda$ and $G$ need to be renormalized whatever scheme is employed.
One finds from (\ref{Ren_3}) the following relations in the case of cutoff regularization
\begin{eqnarray}
\frac{\Lambda^R}{G^R} & = & \frac{\Lambda^B}{G^B}-\frac{1}{4\pi} \left(\Lambda_{\rm UV}^{4}-10\Lambda_{\rm UV}^{2}m^{2}+5m^{4}\log\frac{\Lambda_{\rm UV}^{2}}{m^{2}}\right)\nonumber \\
\frac{1}{G^R} & = & \frac{1}{G^B}+\frac{1}{2\pi} \left(-\frac{23}{3}\Lambda_{\rm UV}^{2}+\frac{13}{3}m^{2}\log\frac{\Lambda_{\rm UV}^{2}}{m^{2}}\right)\,,
\end{eqnarray}
and in the case of dimensional regularization the following
\begin{eqnarray}
\frac{\Lambda^R}{G^R} =  \frac{\Lambda^B}{G^B}-\frac{1}{\epsilon}\frac{5}{2\pi}m^4\qquad\qquad
\frac{1}{G^R} = \frac{1}{G^B}+\frac{1}{\epsilon}\frac{13}{3\pi}m^{2}\,.
\end{eqnarray}
Matter contributions to these relations can be worked out easily using the relative heat kernel coefficients computed in appendix \ref{heat-kernel}.
The beta functions for $\Lambda$ and $G$ are then immediately extracted as minus the coefficient of the poles
\begin{eqnarray}
\mu \partial_{\mu}\! \left(\frac{\Lambda}{G} \right) =  \frac{5}{2\pi}m^4\qquad\qquad
\mu \partial_{\mu}\! \left(\frac{1}{G} \right)  = -\frac{13}{3\pi}m^{2}\,.
\end{eqnarray}
If we now insert $m^2 = -2\Lambda$ we find
\begin{eqnarray}
\mu \partial_{\mu} \Lambda = \frac{4}{3\pi}\Lambda^2 G \qquad\qquad
\mu \partial_{\mu}  G  = -\frac{26}{3\pi}\Lambda G^2\,,
\end{eqnarray}
{ which are the beta functions of the cosmological and Newton's constants within the EFT approach.}
%
%
%

As a final topic we discuss on shell UV divergencies, the only ones that could possibly be physical. For this we need to specify a classical background solution of Einstein's equations. For simplicity we consider the case where no matter is present, so that the EOM becomes just $R_{\mu\nu}=\Lambda g_{\mu\nu}$. In dimensional regularization, the divergent part of the trace (\ref{Ren_3}) then becomes
\begin{equation}
\left.T_{2}\right|_{\rm div} =  -\frac{1}{(4\pi)^{2}}\frac{1}{\epsilon} \int d^{4}x\sqrt{g}\Big[-\frac{58}{5}\Lambda^2 +\frac{53}{45}\textrm{Riem}^2\Big]\,,
\label{Ren_32}
\end{equation}
where we have used $C^2 = \textrm{Riem}^2-\frac{8}{3}\Lambda^2$ to write it in the form as given in \cite{Christensen:1979iy}.
These on--shell divergencies lead to the following beta function for
the cosmological constant%
\begin{equation}
\mu \partial_{\mu}\! \left(\frac{\Lambda}{G} \right) =  \frac{1}{2\pi}\frac{58}{5}\Lambda^2\,,
\end{equation}
which under the assumption that $\Lambda$ is constant \cite{Groh:2013oaa} gives
\begin{equation}
\mu \partial_{\mu} (\Lambda G) =  -\frac{29}{5\pi}(\Lambda G)^2\,,
\end{equation}
so that the dimensionless coupling $\Lambda G$ is asymptotically safe \cite{Fradkin:1981hx}.
Integrating this equation gives a relation similar to (\ref{runR2}).
However other backgrounds have to be considered case by case.

\subsection{Local finite terms}\label{section_finite_local}

The finite physical part of the effective action is independent of the regularization employed
\begin{equation}
T_{\rm finite} = -\frac{1}{2(4\pi)^{d/2}}\sum_{n=\frac{d}{2}+1}\frac{1}{m^{2n-d}}B_{2n}
\underset{d=4}{=} -\frac{1}{2(4\pi)^{2}}\frac{1}{m^{2}}B_{6}+O\!\left(\frac{1}{m^{4}}\right)
 \label{mmm_ea_local}
\end{equation}
which is an expansion in inverse power of the mass. The local
heat kernel coefficients are known, for any kind of matter, to order
$B_{6}$ \cite{Avramidi:2000bm}. 
In particular, for a $\chi=0$ scalar field we find the following finite part
\begin{eqnarray}
T_{\rm finite} & = & -\frac{1}{2(4\pi)^{2}}\frac{1}{m^{2}}\int d^{4}x\sqrt{g}\left[\frac{1}{336}R\square R+\frac{1}{840}R_{\mu\nu}\square R^{\mu\nu}\right.\nonumber \\
&  & +\frac{1}{1296}R^{3}-\frac{1}{1080}RR_{\mu\nu}R^{\mu\nu}-\frac{4}{2835}R_{\mu}^{\nu}R_{\nu}^{\alpha}R_{\alpha}^{\mu}\nonumber \\
&  & +\frac{1}{945}R_{\mu\nu}R_{\alpha\beta}R^{\mu\alpha\nu\beta}+\frac{1}{1080}RR_{\mu\nu\alpha\beta}R^{\mu\nu\alpha\beta}+\frac{1}{7560}R_{\mu\nu}R^{\mu\alpha\beta\gamma}R_{\;\alpha\beta\gamma}^{\nu}\nonumber \\
&  & \left.+\frac{17}{45360}R_{\mu\nu}^{\;\;\alpha\beta}R_{\alpha\beta}^{\;\;\gamma\delta}R_{\gamma\delta}^{\;\;\mu\nu}-\frac{1}{1620}R_{\;\mu\;\nu}^{\alpha\;\beta}R_{\;\gamma\;\delta}^{\mu\;\nu}R_{\;\alpha\;\beta}^{\gamma\;\delta}\right]+O\!\left(\frac{1}{m^{4}}\right)\,.
\label{mmm_ea_local_example}
\end{eqnarray}
The other massive matter cases can be worked out \cite{Avramidi:2000bm,Groh:2011dw} and are given in Table \ref{FLR3}.
As we noted before, the last two terms with three Riemann tensors are not independent in four dimensions.
The inverse mass expansion in (\ref{mmm_ea_local}), when applicable\footnote{Not applicable to gravity even if there is a cosmological constant since the ghosts are in any case massless.},
changes how six, or higher derivative  terms are suppressed in the matter--gravity EFT  $\frac{1}{M^{2n}}\to\frac{1}{m^2M^{2n-2}}$  \cite{Burgess:2003jk}.

General expressions for the $B_{2n}$ are unmanageable, but certain classes of invariants can be
re--summed as we will see in section \ref{section_curvature}.
These expressions, when expanded in the massless limit, give rise to leading logarithmic contributions to which we turn our attention to, in the following section wherein we also discuss their RG improvements.
%
\renewcommand{\arraystretch}{2}
\begin{table}[t]
\begin{centering}
\begin{tabular}{cccccccc}
\hline\hline \\[-1.15cm]
& $0$ & $\frac{1}{2}$      \tabularnewline
\hline 
$R\square R$ & $\frac{1}{336}\left(1-\frac{28}{15}\chi+\frac{7}{9}\chi^{2}\right)$ & $-\frac{1}{560}$  \tabularnewline
\hline 
$R_{\mu\nu}\square R^{\mu\nu}$ & $\frac{1}{840}$ & $\frac{1}{168}$  \tabularnewline
\hline 
$R^{3}$ & $\frac{(1-\chi)^{3}}{1296}$ & $\frac{1}{5184}$   \tabularnewline
\hline 
$RR_{\mu\nu}R^{\mu\nu}$ & $-\frac{1-\chi}{1080}$ & $-\frac{1}{1080}$   \tabularnewline
\hline 
$RR_{\mu\nu\alpha\beta}R^{\mu\nu\alpha\beta}$ & $\frac{1-\chi}{1080}$ & $-\frac{7}{8640}$   \tabularnewline
\hline 
$R^{\mu}_{\;\;\nu}R^{\nu}_{\;\;\alpha}R^{\alpha}_{\;\;\mu}$ & $-\frac{4}{2835}$ & $-\frac{25}{4536}$   \tabularnewline
\hline 
$R_{\mu\nu}R_{\alpha\beta}R^{\mu\alpha\nu\beta}$ & $\frac{1}{945}$ & $-\frac{47}{7560}$   \tabularnewline
\hline 
$R_{\mu\nu}R^{\mu\alpha\beta\gamma}R_{\;\;\alpha\beta\gamma}^{\nu}$ & $\frac{1}{7560}$ & $\frac{19}{7560}$   \tabularnewline
\hline 
$R_{\mu\nu}^{\;\;\;\;\alpha\beta}R_{\alpha\beta}^{\;\;\;\;\gamma\delta}R_{\gamma\delta}^{\;\;\;\;\mu\nu}$ & $\frac{17}{45360}$ & $\frac{29}{45360}$  \tabularnewline
\hline 
$R_{\;\;\mu\;\;\nu}^{\alpha\;\;\beta}R_{\;\;\gamma\;\;\delta}^{\mu\;\;\nu}R_{\;\;\alpha\;\;\beta}^{\gamma\;\;\delta}$ & $-\frac{1}{1620}$ & $-\frac{1}{648}$   \tabularnewline
\hline\hline
\end{tabular}
\par\end{centering}
\caption{Six derivatives invariants present in $I_3$ and their relative $B_6$ heat kernel coefficients for spin $0$ and $\frac{1}{2}$ fields. \label{FLR3}}
\end{table}

\section{Leading logarithmic terms}\label{section_nlna}

Up to now we have examined the local contributions to the effective action coming from the local expansion of the functional trace $T$.
UV divergencies, being local, are the first of these terms, while the other terms lead to finite local corrections when the action is expandable in inverse powers of mass.
When this is not possible, i.e. in the case of massless theories, the leading finite corrections are non--local logarithmic terms in $d=4$ \cite{Barvinsky:1987uw,Barvinsky:1990up,Dalvit:1994gf,Elizalde:1995tx,Gorbar:2002pw,Gorbar:2003yt,Satz:2010uu}.
Being second order in the curvatures, we can write these terms as
\begin{equation}
T_{\rm finite} =  \int d^4x \sqrt{g}\left[ \alpha_{R^2} R \log \frac{-\square}{\mu^2} R +\alpha_{C^2}\, C_{\alpha\beta\gamma\delta}  \log \frac{-\square}{\mu^2} C^{\alpha\beta\gamma\delta}\right] + ...
\label{nlna}
\end{equation}
Both $\alpha_{R^2}$ and $\alpha_{C^2}$ are calculable constants 
with the aid of the non--local heat kernel expansion, as we will show in the next section.
But there is indeed a clever and physically transparent trick that allows the straightforward determination
of these constants.
For dimensional reasons we have introduced the reference scale $\mu$ in (\ref{nlna}),
which indirectly carries the information about the renormalization needed to define the effective action.
Notice that the loop in the trace inevitably leads to terms of the form $\log \frac{-\square}{\Lambda_{\rm UV}^2}$, which is then separated by introducing the scale $\mu$ to obtain the finite terms (\ref{nlna}) and the divergent term $-2\log \frac{\Lambda_{\rm UV}}{\mu}$, or equivalently $-2\frac{1}{\epsilon}$  in dimensional regularization. 
Thus there is a direct link between the coefficients $\alpha_i$ and the coefficients of the one loop beta functions of the couplings of the $R^{2}$ and $C^2$ invariants.
A comparison with (\ref{Ren_5}) then gives
\begin{equation}
\alpha_i=\frac{\gamma_{i}}{2(4\pi)^{2}}\,,
\label{alpha}
\end{equation}
and the coefficients of the logarithmic terms can
be read off from Table \ref{RenR2} without any further computation.
In section \ref{section_curvature} we will check this result with an explicit computation.

\paragraph{A comment on running couplings}

It is precisely the argument exposed in the previous paragraph that explains why and when the RG running with respect to the unphysical parameter $\mu$,
or equivalents, can instead be interpreted as a real, physical running with respect to changes of a physical scale,
as can be the momenta of a photon used to ``look" at a proton.
Take the case of QED. Since the effective action contains a non--local term of the form (\ref{nlna}), 
\begin{equation}
\Gamma_{\rm QED} = \frac{1}{48\pi^{2}}\int d^4x \,F_{\alpha\beta} \log \frac{-\square}{\mu^2} F^{\alpha\beta} + ...\,,
\label{nlnaW}
\end{equation}
then a variation with respect to the physical scale $q$ and the a priori unphysical RG scale $\mu$ are related by
\begin{equation}
2q^2 \partial_{q^2}=-\mu\partial_{\mu}\,.
\label{physrun}
\end{equation}
For this reason the running coupling is straightforwardly well defined in QED or QCD where the couplings are dimensionless.
When couplings are dimensionfull, the effective action will still be physically scale dependent, but one cannot rely on 
a simple relation like (\ref{physrun}) to obtain it and a more careful analysis is indeed needed.

\paragraph{RG improvement}

For dimensional reasons, further corrections\footnote{Not contained in the LO contribution $T$, but in the NLO, NNLO, ... contributions.}
to the leading logarithmic terms in (\ref{nlna}) can be parametrized
as $R\, h_{R^2}(-\square/\mu^{2})\, R$ and equivalently for the Weyl term,
where the $h_i$ are functions of $u=-\square/ \mu^2$.
The logarithmic terms in (\ref{nlna}) can be obtained by solving the following equation,
\begin{equation}
u\partial_{u}h_i(u)= \alpha_i h_i(u)\,,
\label{betah}
\end{equation}
if we set $h_i(u)=1$ on the right hand side of it.
The appearance of the factor $h_i(u)$ on the right hand side of (\ref{betah})
represents instead an RG improvement which, through its solution
gives rise to the log--resummed form
\begin{equation}
h_i(u)=u^{\alpha_i} -1= \alpha_i \log u + \frac{1}{2} (\alpha_i \log u)^2 + ...\,.
\label{logh}
\end{equation}
Using this the RG improved version of  (\ref{nlna}) is then
\begin{equation}
T_{\rm finite} =  \int d^4x \sqrt{g}\left[R \left(\frac{-\square}{\mu^2}\right)^{\alpha_{R^2}}R  + C_{\alpha\beta\gamma\delta} \left(\frac{-\square}{\mu^2}\right)^{\alpha_{C^2}}C^{\alpha\beta\gamma\delta}\right] + ...
\label{nlrg}
\end{equation}
where we note that $0 <\alpha_{i}\ll 1$, even in presence of large numbers of matter fields, and so these corrections are genuinely non--local.
Applications of this class of effective actions have been discussed in \cite{LopezNacir:2006tn}.

\paragraph{Non--local vs non--analytical}

Up to now we discussed non--local terms, but there is also the possibility that the argument of the logarithms is a scalar build out of curvatures, like $R$.
Discarding the Weyl term, we will then have a contribution of the form
\begin{equation}
T_{\rm finite} = \alpha_{R^2} \int d^4x \sqrt{g}\,R \log \frac{R}{\mu^2}  R + ...
\label{nlna2}
\end{equation}
where  $\alpha_{R^2}$ is again given by (\ref{alpha}).
This is a non--analytical contribution which can be found, for example, upon evaluating $T$ on the sphere $S^{4}$,
where the explicit knowledge of the spectrum of the Laplacian allows for a direct computation of the trace--log formula \cite{Avramidi:2000bm}.
The phenomenology of these non--analytical terms is of the $f(R)$ type and has been discussed
extensively, see for example \cite{Alavirad:2013paa,Codello:2014sua,Ben-Dayan:2014isa,Rinaldi:2014gha}.
We here emphasise that such terms are present in the EFT of gravity with definite coefficients.
In particular, the RG improvement can be applied also in this non--analytical case \cite{Codello:2014sua}.
Before we close this section a final question to ask is: when is the action non--analytical and when is it non--local? 
The answer might depend on the properties of the background, in particular compact spaces usually give rise to non--analytical terms while non--compact spaces usually generate non--local terms \cite{Codello:2012kq}. In any case,  both terms can in principle be present in the EFT effective action depending on the context, but probably not together.

\section{Non--local terms via heat kernel}\label{section_curvature}

We can resum a subclass of terms in the series (\ref{mmm_ea_local}) for $T_{\rm finite}$ by
rearranging it and collecting all terms with the same numbers of curvatures.
This naturally leads to a curvature expansion of the form
\begin{equation}
T_{\rm finite} =  \left.T_{\rm finite}\right|_{\mathcal{R}^{2}} +\left.T_{\rm finite}\right|_{\mathcal{R}^{3}} + O(\mathcal{R}^4)
\end{equation}
where each term at a given order can be written in terms of the relative non--local 
heat kernel structure functions, that we shall define in the following.
For example, in equation (\ref{mmm_ea_local_example}) the first term is of the form $R\square R$,
a similar term for $B_8$ will be of the form $R \square^2 R$, and therefore in general in the coefficient
$B_{2n}$ there will be a term of the form $R \square^{n-2}R$.
The series so defined can be resummed using the non--local heat kernel expansion and gives rise to a,
generally non--local, structure function of the variable $\square/m^2$.
Equivalently for the other $\mathcal{R}^{2}$ curvature invariant $R_{\mu\nu} \square R^{\mu\nu}$ and similarly for the other higher order curvature terms that start to appear from $B_8$.
This summation approach allows us to obtain the complete form of $ \left.T_{\rm finite}\right|_{\mathcal{R}^{2}}$ and $\left.T_{\rm finite}\right|_{\mathcal{R}^{3}}$.
The complexity of the latter is very demanding and therefore, in what follows, we will only focus on the curvature squared terms.

\subsection{Curvature expansion to order $\mathcal{R}^{2}$}

Using the non--local heat kernel expansion reported in the appendix \ref{heat-kernel},
we find that the finite part of the curvature square terms in the LO effective action is given by
\begin{equation}
\left.T_{\rm finite}\right|_{\mathcal{R}^{2}}=-\frac{1}{2(4\pi)^{d/2}}\int d^{d}x\sqrt{g}\,\textrm{tr}\,\mathcal{R}\left(\int_{1/\Lambda_{\rm UV}^{2}}^{\infty}\frac{ds}{s}s^{-d/2+2}\left[f_{i}(-s\square)-f_{i}(0)\right]\,e^{-s m^2}\right)\mathcal{R} \,,
\label{sfR2}
\end{equation}
where the $f_i$ for $i=\{Ric,R,RU,U,\Omega\}$ are the non--local heat kernel structure functions given in equation (\ref{sf}).
The subtraction $f_{i}(0)$ reflect exactly the renormalization of the curvature square terms
we performed in section \ref{section_renor} and makes the integrals in the above equation finite in the limit $\Lambda_{\rm UV} \to \infty$. 
This allows us to define the finite $\mathcal{R}^{2}$ structure functions as
\begin{equation}
\gamma_{i}\!\left(\frac{X}{m^{2}}\right)\equiv\lim_{\Lambda_{\rm UV}\to\infty}\int_{1/\Lambda_{\rm UV}^{2}}^{\infty}\frac{ds}{s}s^{-d/2+2}\left[f_{i}(sX)-f_{i}(0)\right]e^{-s m^2}\,.
\label{gamma}
\end{equation}
The great advantage of using this formalism is that the heat kernel structure functions do not depend on $d$ and thus the relation (\ref{gamma}) gives $\left.T_{\rm finite}\right|_{\mathcal{R}^{2}}$ in arbitrary dimensions.

In $d=4$ we find the following form
\begin{eqnarray}
\left.T_{\rm finite}\right|_{\mathcal{R}^{2}} & = & -\frac{1}{2(4\pi)^{2}}\int d^{4}x\sqrt{g}\,\textrm{tr}\left[\mathbf{1}R_{\mu\nu}\gamma_{Ric}\!\!\left(\frac{-\square}{m^{2}}\right)\!R^{\mu\nu}+\mathbf{1}R\,\gamma_{R}\!\!\left(\frac{-\square}{m^{2}}\right)\!R\right.\nonumber \\
 &  & \left.+R\gamma_{RU}\!\!\left(\frac{-\square}{m^{2}}\right)\!\mathbf{U}+\mathbf{U}\gamma_{U}\!\!\left(\frac{-\square}{m^{2}}\right)\!\mathbf{U}+\mathbf{\Omega}_{\mu\nu}\gamma_{\Omega}\!\!\left(\frac{-\square}{m^{2}}\right)\!\mathbf{\Omega}^{\mu\nu}\right]\,,
\label{ea_R2_m}
\end{eqnarray}
where the four dimensional finite $\mathcal{R}^{2}$ structure functions are 
\begin{eqnarray}
\gamma_{Ric}(u) & = & \frac{1}{40}+\frac{1}{12u}-\frac{1}{2}\int_{0}^{1}d\xi\,\left[\frac{1}{u}+\xi(1-\xi)\right]^{2}\log\left[1+u\,\xi(1-\xi)\right]\nonumber \\
\gamma_{R}(u) & = & -\frac{23}{960}-\frac{1}{96u}+\frac{1}{32}\int_{0}^{1}d\xi\,\Bigl\{-1+\frac{2}{u^{2}}+\frac{4}{u}\left[1+\xi(1-\xi)\right]
\nonumber \\ &  & \qquad\qquad\qquad\qquad\qquad\quad\;\;+\,2\xi(2-\xi)(1-\xi^{2})\Bigr\}\log\left[1+u\,\xi(1-\xi)\right]\nonumber \\
\gamma_{RU}(u) & = & \frac{1}{12}-\frac{1}{2}\int_{0}^{1}d\xi\,\left[-\frac{1}{2}+\frac{1}{u}+\xi(1-\xi)\right]\,\log\left[1+u\,\xi(1-\xi)\right]\nonumber \\
\gamma_{U}(u) & = & -\frac{1}{2}\int_{0}^{1}d\xi\,\log\left[1+u\,\xi(1-\xi)\right]\nonumber \\
\gamma_{\Omega}(u) & = & \frac{1}{12}-\frac{1}{2}\int_{0}^{1}d\xi\,\left[\frac{1}{u}+\xi(1-\xi)\right]\,\log\left[1+u\,\xi(1-\xi)\right]\,,
\label{ea_R2_structure}
\end{eqnarray}
and $u\equiv -\square/m^2$.
We remark that these relations give us finite part of the effective action to order $\mathcal{R}^{2}$
for \emph{every} theory whose action's Hessian is a Laplacian of the form $\Delta = -\square \mathbf{1} + \mathbf{U}$.
From these expressions for the structure functions one can understand the behavior of the effective action in the two opposite limits, namely the {\it decoupling} limit $u\ll1$ and the {\it massless} limit $u\gg1$.

\paragraph{Local expansion}

In order to obtain the structure functions in the decoupling limit, we perform a Taylor expansion of (\ref{ea_R2_structure}) around $u=0$ to find
\begin{eqnarray}
\gamma_{Ric}(u) & = & -\frac{u}{840}+\frac{u^{2}}{15120}-\frac{u^{3}}{166320}+O(u^{4})\nonumber \\
\gamma_{R}(u) & = & -\frac{u}{336}+\frac{11u^{2}}{30240}-\frac{19u^{3}}{332640}+O(u^{4})\nonumber \\
\gamma_{RU}(u) & = & \frac{u}{30}-\frac{u^{2}}{280}+\frac{u^{3}}{1890}+O(u^{4})\nonumber \\
\gamma_{U}(u) & = & -\frac{u}{12}+\frac{u^{2}}{120}-\frac{u^{3}}{840}+O(u^{4})\nonumber \\
\gamma_{\Omega}(u) & = & -\frac{u}{120}+\frac{u^{2}}{1680}-\frac{u^{3}}{15120}+O(u^{4})\,.
\label{ea_R2_structure_taylor}
\end{eqnarray}
Note that the linear terms in $u$ of $\gamma_{Ric}$ and $\gamma_{R}$ correctly
reproduce the first two terms of order $\mathcal{R}^{2}$ in equation (\ref{mmm_ea_local_example}).
Similarly, in the case of a scalar with arbitrary $\chi$ or a spinor, one can consistently obtain the results of the first two lines of Table \ref{FLR3}.
The higher order terms in $u$ will lead to the coefficients of the corresponding operators in $B_8, B_{10}$ etc.

\paragraph{Non--local expansion}

In the massless limit we can make a Taylor expansion of the structure functions around $u=\infty$ and obtain\footnote{{While the structure functions (\ref{ea_R2_structure}) are well defined also for negative values of $u$ and thus of $m^2$, the expansion around $u=\infty$ imposes the restriction $m^2>0$ which we will implicitly assume every time we perform the massless limit $u\to\infty$.}}
\begin{eqnarray}
\gamma_{Ric}(u) & = & \frac{23}{450}-\frac{1}{60}\log u+\frac{5}{18u}-\frac{\log u}{6u}+\frac{1}{4u^{2}}-\frac{\log u}{2u^{2}}+O\!\left(\frac{1}{u^{3}}\right)\nonumber \\
\gamma_{R}(u) & = & \frac{1}{1800}-\frac{1}{120}\log u-\frac{2}{9u}+\frac{\log u}{12u}+\frac{1}{8u^{2}}+\frac{\log u}{4u^{2}}+O\!\left(\frac{1}{u^{3}}\right)\nonumber \\
\gamma_{RU}(u) & = & -\frac{5}{18}+\frac{1}{6}\log u+\frac{1}{u}-\frac{1}{2u^{2}}-\frac{\log u}{u^{2}}+O\!\left(\frac{1}{u^{3}}\right)\nonumber \\
\gamma_{U}(u) & = & 1-\frac{1}{2}\log u-\frac{1}{u}-\frac{\log u}{u}-\frac{1}{2u^{2}}+\frac{\log u}{u^{2}}+O\!\left(\frac{1}{u^{3}}\right)\nonumber \\
\gamma_{\Omega}(u) & = & \frac{2}{9}-\frac{1}{12}\log u+\frac{1}{2u}-\frac{\log u}{2u}-\frac{3}{4u^{2}}-\frac{\log u}{2u^{2}}+O\!\left(\frac{1}{u^{3}}\right)\,.
\label{ea_R2_structure_taylor_infty}
\end{eqnarray}
This is indeed the expansion which gives rise to non--local terms in the effective action.
Note that the scheme dependent constant terms can be removed by a finite renormalization and thus we will drop them in the following.
From (\ref{ea_R2_structure_taylor_infty}) we see that in the
strict massless limit, i.e. keeping only the logarithms, the effective action becomes
\begin{eqnarray}
\left.T_{\rm finite}\right|_{\mathcal{R}^{2}} & = & \frac{1}{2(4\pi)^{2}} \int d^{4}x\sqrt{g}\,\textrm{tr}\left[\frac{\mathbf{1}}{60}R_{\mu\nu}\log\frac{-\square}{\mu^{2}}R^{\mu\nu}+\frac{\mathbf{1}}{120}R\log\frac{-\square}{\mu^{2}}R\right.\nonumber \\
 &  & \left.-\frac{1}{6}R\log\frac{-\square}{\mu^{2}}\mathbf{U}+\frac{1}{2}\mathbf{U}\log\frac{-\square}{\mu^{2}}\mathbf{U}+\frac{1}{12}\mathbf{\Omega}_{\mu\nu}\log\frac{-\square}{\mu^{2}}\mathbf{\Omega}^{\mu\nu}\right]\,.
 \label{ea_R2_log}
\end{eqnarray}
It is evident from this action  that these logarithms are indeed the ones predicted in section \ref{section_nlna}.
To be able to check equation (\ref{alpha}) we first need to write (\ref{ea_R2_log}) in the Weyl basis. 

\paragraph{Weyl basis}

In $d=4$ the shift to the Weyl basis is defined by
\begin{eqnarray}
\gamma_{C}(u) & = & \frac{1}{2}\gamma_{Ric}(u)\nonumber \\
\gamma_{Rbis}(u) & = & \frac{1}{3}\gamma_{Ric}(u)+\gamma_{R}(u)\,,
\label{ea_R2_structure_Weyl}
\end{eqnarray}
while the other structure functions remain unchanged. The shift between two basis is made using the generalized Euler identity \cite{Avramidi:2000bm,Codello:2012kq}, which also shows that the difference between the structure functions in different basis is of order $\mathcal{R}^3$.
In the Weyl basis the heat kernel structure functions depend on $d$ since both the definitions
of the Weyl tensor and the Euler invariant contain it while in the Ricci basis they do not.
For this reason up to now we have been employing the Ricci basis.

The local and non--local expansions for $\gamma_{C}$ and $\gamma_{Rbis}$ are
\begin{eqnarray}
\gamma_{C}(u) & = & -\frac{u}{1680}+\frac{u^{2}}{30240}-\frac{u^{3}}{332640}+O(u^{4})\nonumber \\
\gamma_{Rbis}(u) & = & -\frac{17u}{5040}+\frac{u^{2}}{2592}-\frac{59u^{3}}{997920}+O(u^{4})\label{ea_R2_structure_Weyl_taylor-1}
\end{eqnarray}
and
\begin{eqnarray}
\gamma_{C}(u) & = & \frac{23}{900}-\frac{1}{120}\log u+\frac{5}{36u}-\frac{\log u}{12u}+\frac{1}{8u^{2}}-\frac{\log u}{4u^{2}}+O\!\left(\frac{1}{u^{3}}\right)\nonumber \\
\gamma_{Rbis}(u) & = & \frac{19}{1080}-\frac{1}{72}\log u-\frac{7}{54u}+\frac{\log u}{36u}+\frac{5}{24u^{2}}+\frac{\log u}{12u^{2}}+O\!\left(\frac{1}{u^{3}}\right)\,.\label{ea_R2_structure_Weyl_taylor_infty-1}
\end{eqnarray}
In the Weyl basis, the effective action (\ref{ea_R2_log}) in the massless limit  therefore becomes
\begin{eqnarray}
\left.T_{\rm finite}\right|_{\mathcal{R}^{2}} & = & \frac{1}{2(4\pi)^{2}} \int d^{4}x\sqrt{g}\,\textrm{tr}\left[\frac{\mathbf{1}}{120}C_{\mu\nu\alpha\beta}\log\frac{-\square}{\mu^{2}}C^{\mu\nu\alpha\beta}+\frac{\mathbf{1}}{72}R\log\frac{-\square}{\mu^{2}}R\right.\nonumber \\
 &  & \left.-\frac{1}{6}R\log\frac{-\square}{\mu^{2}}\mathbf{U}+\frac{1}{2}\mathbf{U}\log\frac{-\square}{\mu^{2}}\mathbf{U}+\frac{1}{12}\mathbf{\Omega}_{\mu\nu}\log\frac{-\square}{\mu^{2}}\mathbf{\Omega}^{\mu\nu}\right]\,.
 \label{ea_R2_log_W}
\end{eqnarray}
We can now explicitly check the validity of (\ref{alpha}) and therefore of the arguments presented in section \ref{section_nlna}.

\subsection{Effective action in $d=2$}

It is useful to check our relations in a case where the explicit form of the LO effective action is exactly known: the Polyakov action.
In $d=2$ we have $R_{\mu\nu}=\frac{1}{2}g_{\mu\nu}R$, or equivalently $C_{\mu\nu\alpha\beta}=0$, and thus
there is only one gravitational structure function
\begin{equation}
\gamma_{R2d}(u)=\frac{1}{2}\gamma_{Ric}(u)+\gamma_{R}(u)\,.
\end{equation}
The curvature square part of the effective action then is
\begin{equation}
\left.\Gamma_{\rm finite}\right|_{\mathcal{R}^{2}}=-\frac{1}{8\pi \,m^2}\int d^{2}x\sqrt{g}\, R\gamma_{R2d}\!\left(\frac{-\square}{m^{2}}\right)\!R\,.
\label{ea_R2_2d}
\end{equation}
For the case of a minimally coupled massive scalar, employing equation (\ref{gamma}),  we find
\begin{equation}
\gamma_{R2d}(u)=\frac{1}{12u}-\frac{1}{u^2}+\frac{2\tanh^{-1}\sqrt{\frac{u}{u+4}}}{u\sqrt{u^3(u+4)}}\,,
\label{polmass}
\end{equation}
with the following expansions around $u=0$ and $u=\infty$ 
\begin{eqnarray}
\gamma_{R2d}(u) & = & \frac{1}{60}-\frac{u}{280}+\frac{u^{2}}{1260}+O(u^{3})\nonumber \\
\gamma_{R2d}(u) & = & \frac{1}{12u}-\frac{1}{2u^2}+\frac{\log u}{u^{3}}+O\!\left(\frac{1}{u^{4}}\right)\,.
\label{ea_R2_taylor}
\end{eqnarray}
Similar expressions can be easily obtained for the fermions and gauge fields.
Gravitons do not propagate in $d=2$, so there is no contribution from the Einstein--Hilbert action. 
In the massless limit $u\rightarrow\infty$, the last equation gives
$\gamma_{R2d}(\infty)=\frac{1}{12u}$ and we correctly recover the Polyakov action \cite{Polyakov:1981rd}
\begin{equation}
\left.\Gamma_{\rm finite}\right|_{\mathcal{R}^{2}}=-\frac{1}{96\pi}\int d^{2}x\sqrt{g}\, R\frac{1}{-\square}R\,.
\label{polyakov}
\end{equation}
Note that only in the massless limit, the $\mathcal{R}^{3}$ or the higher order terms vanish.
On the other hand, the massive version of the Polyakov action to order $\mathcal{R}^2$ is obtained by combining equation (\ref{ea_R2_2d}) together with  equation (\ref{polmass}).


\subsection{LO effective action to order $\mathcal{R}^2$}

We can finally combine our findings for the local and non--local terms and
write down the effective action to LO in the Weyl basis as
\begin{eqnarray}
\Gamma[g] & = & \frac{1}{16\pi G}\int d^{4}x\sqrt{g}\left(2\Lambda-R\right)+\frac{1}{2\lambda}\int d^{4}x\sqrt{g}\, C^{2}+\frac{1}{\xi}\int d^{4}x\sqrt{g}\, R^{2}\nonumber \\
 &&+\int d^{4}x\sqrt{g}\, C_{\mu\nu\alpha\beta}\, \mathcal{G}\!\left(\frac{-\square}{m^2}\right)C^{\mu\nu\alpha\beta}+\int d^{4}x\sqrt{g}\, R\, \mathcal{F}\!\left(\frac{-\square}{m^2}\right)R+O(\mathcal{R}^{3})\,.
\label{ea_R2_Final}
\end{eqnarray}
Here $G,\Lambda,\xi$ and $\lambda$ are the renormalized couplings or phenomenological parameters, i.e. those measured in experiments or observations.
Thus the first line of (\ref{ea_R2_Final}) is the {\it input} to the EFT of gravity
while the second line instead represents the {\it output}, i.e. the universal prediction common to all possible UV completions.
Note that in (\ref{ea_R2_Final}) in order to write down the Einstein--Hilbert part in the conventional form we have rescaled the effective action as $\Gamma \to 16\pi G\, \Gamma$.
The final non--local structure functions $\mathcal{G}$ and $\mathcal{F}$ are,
in the cases of scalars, fermions, abelian gauge fields, given by
\begin{eqnarray}
\mathcal{G}_{0}(u) & = &  -\frac{1}{2(4\pi)^{2}} \left( \frac{1}{2}\gamma_{Ric}(u) \right)\nonumber\\
\mathcal{F}_{0}(u) & = & -\frac{1}{2(4\pi)^{2}} \left(  \frac{1}{3}\gamma_{Ric}(u)+\gamma_{R}(u)+\frac{\chi}{6} \gamma_{RU}(u)+\frac{\chi^2}{36} \gamma_{U}(u) \right)\nonumber\\
\nonumber\\
\mathcal{G}_{\frac{1}{2}}(u) & = &-\frac{1}{2(4\pi)^{2}} \Big(\!-2\gamma_{Ric}(u)+\gamma_{\Omega}(u) \Big)\nonumber\\
\mathcal{F}_{\frac{1}{2}}(u) & = & -\frac{1}{2(4\pi)^{2}} \left(-\frac{4}{3}\gamma_{Ric}(u)-4\gamma_{R}(u)-\gamma_{RU}(u)-\frac{1}{4} \gamma_{U}(u)+\frac{1}{6}\gamma_{\Omega}(u) \right)\nonumber\\
\nonumber\\
\mathcal{G}_{1}(u) & = & -\frac{1}{2(4\pi)^{2}} \left( \gamma_{Ric}(u)+\frac{1}{2}\gamma_{U}(u)-2\gamma_{\Omega}(u) \right)\nonumber\\
\mathcal{F}_{1}(u) & = & -\frac{1}{2(4\pi)^{2}} \left( \frac{2}{3}\gamma_{Ric}(u)+2\gamma_{R}(u)+\gamma_{RU}(u)+\frac{1}{3} \gamma_{U}(u)-\frac{1}{3}\gamma_{\Omega}(u)\right)\,,
\label{strmatter}
\end{eqnarray}
while in the case of gravity they are
\begin{eqnarray}
\mathcal{G}_{2}(u) & = &-\frac{1}{2(4\pi)^{2}} \Big( 5\gamma_{Ric}(u)+3\gamma_{U}(u)-12 \gamma_{\Omega}(u)   \nonumber\\
&&  \qquad\qquad\qquad-4\gamma_{Ric}(u)-\gamma_{U}(u)+4 \gamma_{\Omega}(u) \Big)\nonumber\\
\mathcal{F}_{2}(u) & = &-\frac{1}{2(4\pi)^{2}} \left( \frac{10}{3}\gamma_{Ric}(u)+10\gamma_{R}(u)+6\gamma_{RU}(u)+4\gamma_{U}(u)-2\gamma_{\Omega}(u)\right.\nonumber\\
 &&\left.  \qquad\qquad\qquad-\frac{8}{3}\gamma_{Ric}(u)-8\gamma_{R}(u)+2\gamma_{RU}(u)-\frac{2}{3}\gamma_{U}(u)+\frac{2}{3}\gamma_{\Omega}(u)\right)\,,
\label{strgravity}
\end{eqnarray}
where the first line in this last equation is the graviton contribution with {$m^2= -2\Lambda+V(v)/M^2$}
 while the second line corresponds to the ghost contribution with $m^2=\mu^2$.

\begin{figure}[!t]
\begin{center}
\includegraphics[scale=0.78]{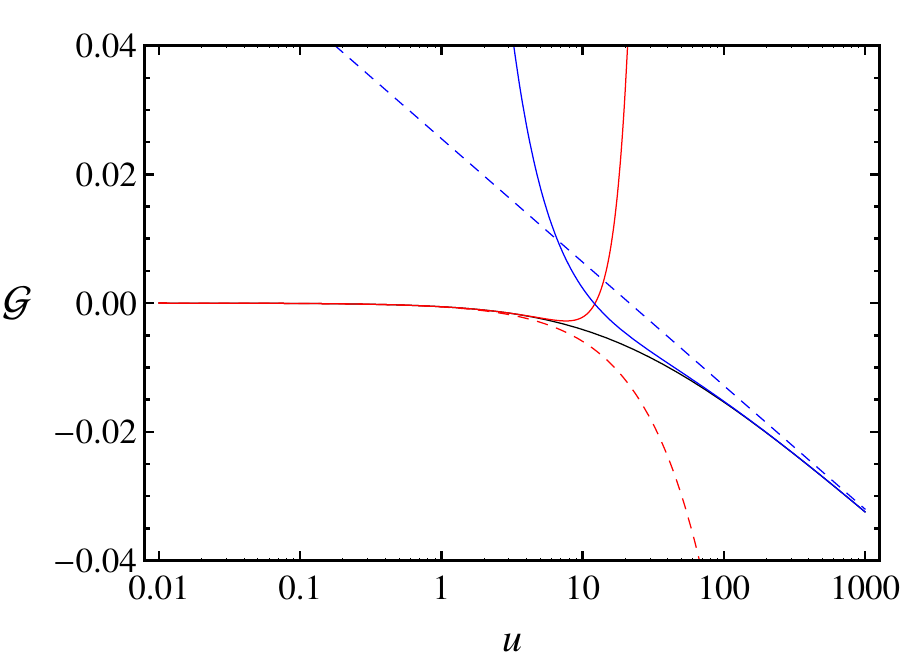}\hskip 20pt
\includegraphics[scale=0.81]{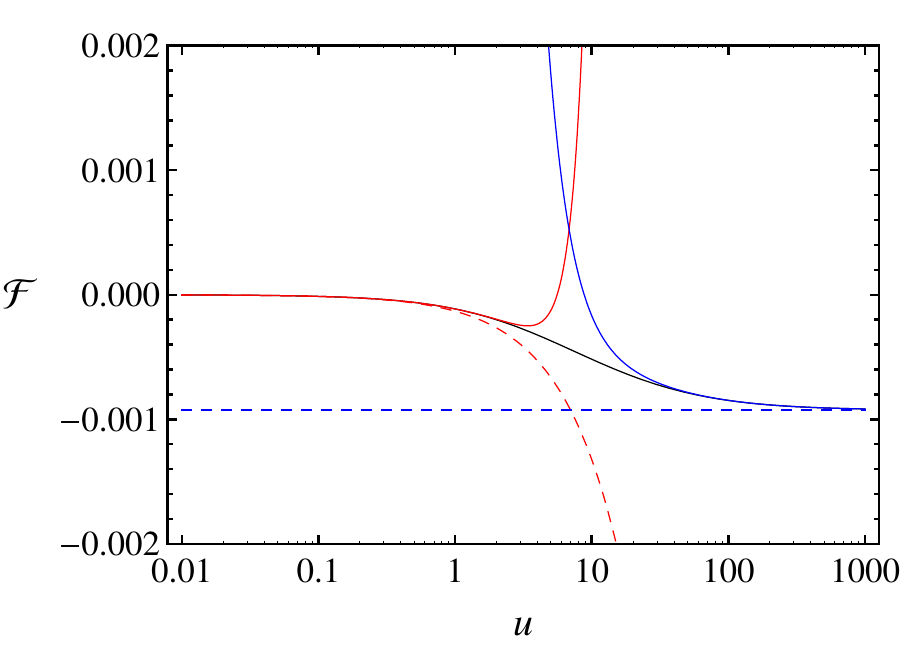}
\end{center}
\caption{Non--local structure functions {(black)} $\mathcal{G}$ and $\mathcal{F}$ for a conformal scalar together with their small $u$ {(red)} and large $u$ {(blue)} approximations.
The dashed lines represent the leading contributions while the solid lines indicate the contributions of the first four terms in each case.
Note that the small $u$ expansion is much less sensitive to subleading corrections than the large $u$ counterpart.
}
\label{FGPlots}
\end{figure}

To make explicit the non--local terms contained in the effective action (\ref{ea_R2_Final}) we expand (\ref{strmatter}) and (\ref{strgravity}) in the massless limit around $u=\infty$.
For example in the case of a minimally coupled scalar ($\chi=0$) we find
\begin{eqnarray}
\mathcal{G}_{0}\!\left(\frac{-\square}{m^2}\right) & = &\frac{1}{(4\pi)^{2}}   \left( \frac{1}{240} \log\!\frac{-\square}{m^{2}} - \frac{5}{72} \frac{m^{2}}{-\square} +\frac{1}{24}\frac{m^{2}}{-\square}\!\log\!\frac{-\square}{m^{2}}-\frac{1}{16}\frac{m^{4}}{\square^{2}}+...\right) \nonumber\\\
\mathcal{F}_{0}\!\left(\frac{-\square}{m^2}\right) & = &\frac{1}{(4\pi)^{2}} \left( \frac{1}{144} \log\!\frac{-\square}{m^{2}} + \frac{7}{108} \frac{m^{2}}{-\square} -\frac{1}{72}\frac{m^{2}}{-\square}\!\log\!\frac{-\square}{m^{2}}-\frac{15}{144}\frac{m^{4}}{\square^{2}}+...\right)\,,
\end{eqnarray}
while the coefficients for the other cases are reported in Table \ref{NLHK}.
Note that the conformal invariant matter has vanishing leading logarithms in the Ricci scalar sector as expected.
Moreover, the parameter $\chi$ also drops out of the Weyl tensor sector.
%
\renewcommand{\arraystretch}{2}
\begin{table}[t]
\begin{centering}
\begin{tabular}{cccca}
\hline\hline \\[-1.15cm]
\mc{1}{}  & \mc{1}{0} & \mc{1}{$\frac{1}{2}$} & \mc{1}{1} & \mc{1}{2} \tabularnewline
\hline 
$m^{2}$ & $m_{\phi}^{2}=V''(v)$ & $m_{\psi}^{2}$ & $\mu^{2}$ & ${-2\Lambda+V(v)/M^2}$  \tabularnewline
\hline 
$C\log\!\frac{-\square}{m^{2}}C$ & $\frac{1}{240}$ & $\frac{1}{40}$ & $\frac{1}{20}$ & $\frac{7}{40}$ \tabularnewline
\hline 
$C\frac{m^{2}}{-\square}C$ & $-\frac{5}{72}$ & $\frac{1}{36}$ & $\frac{11}{18}$ & $\frac{137}{36}$ \tabularnewline
\hline 
$C\frac{m^{2}}{-\square}\!\log\!\frac{-\square}{m^{2}}C$ & $\frac{1}{24}$ & $\frac{1}{12}$ & $-\frac{1}{6}$ & $-\frac{13}{12}$ \tabularnewline
\hline 
$C\frac{m^{4}}{\square^{2}}C$ & $-\frac{1}{16}$ & $\frac{5}{8}$ & $-\frac{3}{4}$ & $-\frac{35}{8}$\tabularnewline
\hline 
$R\log\!\frac{-\square}{m^{2}}R$ & $\frac{(1-\chi)^2}{144}$ & $0$ & $0$ & $\frac{1}{8}$\tabularnewline
\hline 
$R\frac{m^{2}}{-\square}R$ & $\frac{14-18 \chi+3 \chi ^2}{216}$ & $\frac{2}{27}$ & $-\frac{13}{108}$ & $\frac{4}{27}$\tabularnewline
\hline 
$R\frac{m^{2}}{-\square}\!\log\!\frac{-\square}{m^{2}}R$ & $-\frac{1-\chi ^2}{72}$ & $-\frac{1}{36}$ & $\frac{1}{18}$ & $\frac{49}{36}$ \tabularnewline
\hline 
$R\frac{m^{4}}{\square^{2}}R$ & $-\frac{15-6 \chi-\chi ^2}{144}$ & $\frac{1}{6}$ & $0$ & $\frac{17}{24}$\tabularnewline\hline\hline
\end{tabular}
\par\end{centering}
\caption{Non--local coefficients for the various spins.
All numbers should be multiplied by $\frac{1}{(4 \pi)^2}$. 
\label{NLHK}}
\end{table}
In the last grey shaded column of Table \ref{NLHK}, only the leading logarithms receive contributions from both the graviton and the ghosts,
while the other terms, i.e. those which vanish in the massless limit, get only graviton contributions.
It is also understood that the mass in the leading logarithms can be, modulo finite renormalizations, either {$-2\Lambda+V(v)/M^2$} or $\mu^2$.
{The mass in the other logarithms 
and in the overall mass terms are to be interpreted as $-2\Lambda+V(v)/M^2$.}
In Figure \ref{FGPlots}, we have plotted the structure functions $\mathcal{G}$ and $\mathcal{F}$ of a scalar together with their small and large $u$ approximations.
From this example we clearly see that the knowledge of the two expansions allows us to determine the threshold value of $u$ which separates the decoupling from the massless limit, which is a priori not obvious. It also guides us in determining the terms which are relevant in a given regime, especially when solving the effective EOM.

The action (\ref{ea_R2_Final}) is derived in the Euclidean signature while in order to understand the imprints in a physical context, we need to switch it to the Lorentzian signature. 
The final action is then
\begin{eqnarray}
\!\!\!\!\!\!\!\!  \Gamma[g] & = & \frac{1}{16\pi G}\int d^{4}x\sqrt{-g}\left(R-2\Lambda \right)-\frac{1}{2\lambda}\int d^{4}x\sqrt{-g}\, C^{2}-\frac{1}{\xi}\int d^{4}x\sqrt{-g}\, R^{2}\nonumber \\
 &&-\int d^{4}x\sqrt{-g}\, C_{\mu\nu\alpha\beta}\, \mathcal{G}\!\left(\frac{-\square}{m^2}\right)C^{\mu\nu\alpha\beta}-\int d^{4}x\sqrt{-g}\, R\, \mathcal{F}\!\left(\frac{-\square}{m^2}\right)\!R+O(\mathcal{R}^{3})\,,
\label{ea_R2_Final_L}
\end{eqnarray}
where evidently the operator $\square$ is constructed with the Lorentzian metric and more importantly, the Green's function $\frac{1}{-\square}$ is to be interpreted as the retarded Green's function to preserve causality \cite{Barvinsky:1987uw,Vilkovisky:1992za}. To conclude, the action (\ref{ea_R2_Final_L}) is the effective action including  the LO quantum gravitational corrections upto the second order in the curvatures, we should do physics with and it can be applied to any arbitrary background.  
%

\section{Discussion and Outlook}\label{conclusions}


In this paper, we have developed the EFT of gravity in a completely covariant formalism as a saddle point, or a loop expansion in the inverse powers of the Planck mass scale. We have considered both the cases of pure gravity (with the cosmological constant) as well as gravity coupled to matter. By doing so, we have identified the classical theory together with the LO and NLO corrections and also outlined the structure of the NNLO terms in the effective action. We particularly focussed our attention on the local and non--local correction terms at the LO which were computed using the heat kernel methods. In this process, we also discussed some subtleties associated with the regularization and renormalization of the effective action and finally computed the finite part of the effective action expanded to the second order in the curvatures. We found that the LO local terms in our finite effective action, consist of the $R^2$ and $C^2$ operators with their coefficients being phenomenological parameters fixed by observations while in the non--local sector, the various terms, to the second order in the curvatures, appear with their respective structure functions which are completely determined by the covariant EFT of gravity. 

We would like to stress that since the final effective action in our EFT of gravity has been computed in a complete covariant manner, it can be readily used on an arbitrary background and therefore, can be employed to understand different phenomenology. In particular, we have shown that there exist LO quantum gravitational effects which are less suppressed than the well known corrections to Newton's potential but these effects are only present on a spacetime with non--zero curvature and cannot be observed on Minkowski spacetime. Indeed, the quantum corrections to Newton's potential are of order $\hbar/M^2$ and, when compared to the LO quantum corrections of order $\hbar$, are suppressed by an additional factor of $1/M^2$. This fact is one main reason to study and understand in depth the physical implications of the LO corrections and is also the primary reason behind developing the covariant EFT of gravity.

Our formalism allows us to consider the inclusion of the cosmological constant in the framework of the EFT of gravity as Minkowski spacetime is not a solution to the Einstein EOM in the presence of a cosmological constant and our ability of quantize gravity in an arbitrary background comes as a rescue to this. We emphasize that this is indeed an important step since observations indicate the presence of a non--zero cosmological constant and, even in the absence of matter, its presence induces all the non--local terms in the effective action which, in principle, could have a physically observable effect.

The non--local terms that we have shown to appear in the effective action have coefficients which are completely determined by the EFT of gravity and thus, they are indeed a distinct prediction of our formalism. Some of these non--local terms have been recently considered in the literature as consistent non--local modifications of GR and have been studied in the cosmological context as for example candidates of dark energy in order to explain the current acceleration of the universe  \cite{Espriu:2005qn,Deser:2007jk,Koivisto:2008xfa,Biswas:2010zk,Nojiri:2010pw,Barvinsky:2011hd,Barvinsky:2011rk,Solodukhin:2012ar,Maggiore:2013mea,Ferreira:2013tqn,Maggiore:2014sia,Maggiore:2015rma}, in the context of black hole solutions \cite{Bronnikov:2009az,Kehagias:2014sda,Dirian:2014xoa} and also in the Newtonian limit \cite{Koivisto:2008dh,Conroy:2014eja}. Non--local modifications of GR appear also in the context of super--renormalizable theories \cite{Modesto:2011kw,Modesto:2014lga}. So far, these terms have been studied on their own together with the Einstein--Hilbert action but we have clearly shown that all such non--local terms appear together in the EFT of gravity and therefore, should all be included in order to construct a viable scenario of the universe. Furthermore, we have also understood that non--local terms are present in the massless limit, where the effective mass is much smaller than the energy scale of the process. On the contrary, in the decoupling limit, only local terms are present in the effective action.

To gauge the magnitude of LO corrections in a curved background, we will specialize the effective action to the FRW spacetime, derive the effective Friedmann EOM and study the corresponding solutions, in a follow up paper \cite{Codello:2015pga}. 
In future, we also plan to study black hole solutions for the effective action of the EFT of gravity and successively understand the role of conformal anomalies that are not considered in this paper as we restrict our analysis to curvature squared terms.
We also intend to derive the covariant form of the effective action corresponding to the matter--gravity EFT, in order to obtain the known corrections to Newton's potential.

To conclude, the EFT formalism is a consistent approach which is able to access the low energy phenomenology of quantum gravity but it does not solve any of the fundamental  problems associated with it, as it breaks down before the Planck scale. In any case, a UV completion of gravity capable of resolving these issues will lead to the same low energy phenomenology as the EFT and it may very well be, that the first quantum gravity imprints that we will ever observe are already described by the EFT formalism.

\paragraph*{Acknowledgments}

A.C. would like to thank John F. Donoghue, Basem Kamal El--Menoufi and Roberto Percacci for useful discussions.
R.K.J. acknowledges financial support from the Danish Council for Independent Research in Natural Sciences for part of the work.
The CP$^3$-Origins centre is partially funded by the Danish National Research Foundation, grant number DNRF90.

\appendix

\section{Heat kernel methods}\label{heat-kernel}

When working on an arbitrary background one needs a covariant way to compute the loop diagrams.
A powerful set of techniques to perform this task is based on the heat kernel and its expansion \cite{Vilkovisky:1992za,Avramidi:2000bm,Vassilevich:2003xt}.
When dealing with one loop diagrams, as those entering the LO corrections,
one has to evaluate only the heat kernel trace, fact that facilitates computations.
The heat kernel $K_{xy}(s)$ for a Laplace type differential operator $\Delta= -\square\,{\bf 1} + {\bf U}$ is defined by
\begin{equation}
(\partial_{s}+\Delta_{x})K_{xy}(s)=0\qquad\qquad K_{xy}(0)=\delta_{xy}\,.
\label{HK_3-1}
\end{equation}
Formally we can write the solution as $K_{xy}(s)=e^{-s\Delta_{x}}\delta_{xy}$ and define its trace as $K(s)=\textrm{Tr}\,e^{-s\Delta}$.
One can thus write the trace--log as a parametric integral in $s$ over
the heat kernel trace directly converting any expansion of the heat kernel in an expansion for the one loop diagram.

Unfortunately, no exact expressions for the heat kernel or its trace are known  on an arbitrary background and so we have to resort to 
asymptotic expansions, which in any case are very useful.
There are two possible expansions for the heat kernel, both of them have been used in this paper: the local expansion, which is an expansion in the derivatives and the non--local expansion, which is an expansion in the (generalized) curvatures, which we review in the following.

\subsection{Local heat kernel expansion}

The local expansion of the heat kernel trace is
\begin{equation}
K(s) = \frac{1}{(4\pi)^{d/2}}\sum_{n=0}^{\infty}s^{n-d/2}B_{2n}(\Delta)\,,
\label{HK_4}
\end{equation}
where the integrated heat kernel coefficients $B_{2n}(\Delta)$ are related to the unintegrated coefficients $b_{2n}(\Delta)$ by
\begin{equation}
B_{2n}(\Delta)=\int d^{d}x\,\textrm{tr}\, b_{2n}(\Delta)\,.
\label{HK_5}
\end{equation}
The most useful fact about the unintegrated heat kernel coefficients is that they do not depend on the number of dimensions. 
The first three unintegrated coefficients are
\begin{eqnarray}
b_{0}(\Delta) & = & \mathbf{1}\nonumber \\
b_{2}(\Delta) & = & \mathbf{1}\frac{R}{6}-\mathbf{U}\nonumber \\
b_{4}(\Delta) & = & \frac{1}{2}\mathbf{U}^{2}-\frac{1}{6}R\mathbf{U}+\frac{1}{6}\square\mathbf{U}+\frac{1}{12}\mathbf{\Omega}_{\mu\nu}\mathbf{\Omega}^{\mu\nu}\nonumber \\
&  & +\,\mathbf{1}\!\left(\frac{1}{180}R_{\mu\nu\alpha\beta}^{2}-\frac{1}{180}R_{\mu\nu}^{2}+\frac{1}{72}R^{2}-\frac{1}{30}\square R\right)\,,
\label{HK_6}
\end{eqnarray}
where $\mathbf{\Omega}_{\mu\nu}\equiv[\nabla_{\mu},\nabla_{\nu}]$.
The next, $b_{6}(\Delta)$ starts already to be quite cumbersome \cite{Avramidi:2000bm,Vassilevich:2003xt};
for a recent computation see \cite{Groh:2011dw}.
The coefficient $b_{8}(\Delta)$ is known only in special cases. 

Using the traces reported in Table \ref{reltrace} we can evaluate explicitly the heat kernel coefficients (\ref{HK_6}) in $d=4$ for the various Laplacians considered in the paper.
By means of the following relations linking the different basis,
\begin{eqnarray}
\frac{1}{180}\textrm{Riem}^{2}-\frac{1}{180}\textrm{Ric}^{2}+\frac{1}{72}R^{2} =  \frac{1}{60}\textrm{Ric}^{2}+\frac{1}{120}R^{2}+\frac{1}{180}E
= \frac{1}{120}C^{2}+\frac{1}{72}R^{2}-\frac{1}{360}E\,,
\label{R2_2}
\end{eqnarray}
we can write an expression for $B_{4}(\Delta)$ as
\begin{equation}
B_{4}(\Delta) =  \int d^{4}x\sqrt{g}\big(\gamma_{C^2}\,C^{2}+\gamma_{R^2}\,R^{2}+\gamma_{E}\,E+\gamma_{\square R}\,\square R\big)\,,
\label{B4}
\end{equation}
where the gamma coefficients are reported in Table \ref{RenR2}. Obviously, for photons and gravitons the combinations $B_4(\Delta_1)-2B_4(-\square)$ and $B_4(\Delta_2)-2B_4(\Delta_{gh})$ have been considered, while for the fermions a minus sign has been added.
The heat kernel coefficients $B_0(\Delta)$ and $B_2(\Delta)$ are easily obtained from  Table \ref{reltrace}.

\renewcommand{\arraystretch}{2}
\begin{table}[t]
\begin{centering}
\begin{tabular}{ccccc}
\hline\hline \\[-1.15cm]
\mc{1}{}  & \mc{1}{$\textrm{tr}\,\mathbf{1}$} & \mc{1}{$\textrm{tr}\,\mathbf{U}$} & \mc{1}{$\textrm{tr}\,\mathbf{U}^2$} & \mc{1}{$\textrm{tr}\,\mathbf{\Omega}_{\mu\nu}\mathbf{\Omega}^{\mu\nu}$} \tabularnewline
\hline 
$\Delta_0$ & $1$ & $\frac{\chi}{6}R$ & $\frac{\chi^2}{36}R^2$ & $0$ \tabularnewline
\hline 
$\Delta_{\frac{1}{2}}$ & $4$ & $R$ & $\frac{1}{4}R^2$ & $-\frac{1}{2}\textrm{Riem}^{2}$ \tabularnewline
\hline 
$\Delta_{1}$ & $4$ & $R$ & $\textrm{Ric}^{2}$ & $-\textrm{Riem}^{2}$ \tabularnewline
\hline 
$\Delta_{2}$ & $10$ & $6R$ & $5R^{2}-6\,\textrm{Ric}^{2}+3\,\textrm{Riem}^{2}$ & $-6\,\textrm{Riem}^{2}$\tabularnewline
\hline 
$\Delta_{gh}$ & $4$ & $-R$ & $\textrm{Ric}^{2}$ & $-\textrm{Riem}^{2}$\tabularnewline
\hline\hline
\end{tabular}
\par\end{centering}
\caption{Relevant traces needed for the computation of the heat kernel coefficients $B_0, B_2$ and $B_4$ in $d=4$.
These traces also offer the expressions needed to evaluate the non--local heat kernel trace $K_{\mathcal{R}^2}(s)$. \label{reltrace}}
\end{table}

\subsection{Non--local heat kernel expansion}

In order to calculate the finite non--local parts of the effective action we need
a more sophisticated version of the heat kernel expansion which resums
an infinite number of local heat kernel coefficients. This expansion has
been developed in \cite{Barvinsky:1987uw,Barvinsky:1990up,Barvinsky:1990uq,Barvinsky:1993en,Avramidi:2000bm} and retains the infinite number of heat
kernel coefficients in the form of non--local structure functions. Keeping terms up to second order in the (generalized) curvatures,
 the non--local heat kernel expansion is
\begin{equation}
K(s) = K_{\mathcal{R}^{0}}(s)+K_{\mathcal{R}}(s)+K_{\mathcal{R}^{2}}(s)+O(\mathcal{R}^{3})
\end{equation}
where $K_{\mathcal{R}^{0}}(s)$ and $K_{\mathcal{R}}(s)$ are just the $n=0,1$ terms of the local expansion (\ref{HK_4}), while
\begin{eqnarray}
K_{\mathcal{R}^{2}}(s) & = & \frac{s^{2}}{(4\pi s)^{d/2}}\int d^{d}x\sqrt{g}\,\textrm{tr}\bigg\lbrace \mathbf{1}R_{\mu\nu}f_{Ric}(-s\square)R^{\mu\nu}+\mathbf{1}R\, f_{R}(-s\square)R\nonumber\\
&  & \qquad\qquad\qquad+\, R\, f_{RU}(-s\square)\mathbf{U}+\mathbf{U}f_{U}(-s\square)\mathbf{U}
+\,\mathbf{\Omega}_{\mu\nu}f_{\Omega}(-s\square)\mathbf{\Omega}^{\mu\nu}\bigg\rbrace \,.
\label{nlhk}
\end{eqnarray}
The non--local heat kernel structure functions entering this expression are\footnote{We note that the authors in \cite{Barvinsky:1990up} use a slightly different basis where $U=-P+\frac{R}{6}$.}
\begin{eqnarray}
f_{Ric}(x) & = & \frac{1}{6x}+\frac{1}{x^{2}}\left[f(x)-1\right]\nonumber \\
f_{R}(x) & = & \frac{1}{32}f(x)+\frac{1}{8x}f(x)-\frac{7}{48x}-\frac{1}{8x^{2}}\left[f(x)-1\right]\nonumber \\
f_{RU}(x) & = & -\frac{1}{4}f(x)-\frac{1}{2x}\left[f(x)-1\right]\nonumber \\
f_{U}(x) & = & \frac{1}{2}f(x)\nonumber \\
f_{\Omega}(x) & = & -\frac{1}{2x}\left[f(x)-1\right]\,,
\label{sf}
\end{eqnarray}
where the basic non--local heat kernel structure function $f(x)$ appearing in these relations is defined
in terms of the following parametric integral
\begin{equation}
f(x)=\int_{0}^{1}d\xi\, e^{-x\xi(1-\xi)}\,.
\label{bsf}
\end{equation}
A simple derivation of the structure functions (\ref{sf}) using Feynman diagrams has been given in \cite{Codello:2012kq}.
The Ricci basis non--local heat kernel structure functions (\ref{sf}) are independent of the number of spacetime dimensions, as are the local heat kernel coefficients of which they are the resummation.
They can be used to compute all one loop bubble diagrams in any dimension for any theory in which the Hessian is a Laplace type operator, and probably represent the most efficient way to perform such computations.
If we Taylor expand the non--local structure functions (\ref{sf}) and insert them in (\ref{nlhk}),
they will reproduce the local heat kernel coefficients of the invariants with two curvatures,
modulo total derivatives \cite{Barvinsky:1990up,Codello:2012kq}.

\bibliographystyle{JHEP}
\bibliography{EFT_GR_I_Bibliography}

\end{document}